\documentclass[useAMS,usenatbib]{mn2e}

\usepackage{graphicx}
\usepackage{txfonts}
\usepackage[authoryear]{natbib}
\usepackage{subfigure}

\title[Detecting planets using eclipse timing]{Detecting circumbinary planets using eclipse timing of binary 
stars - numerical simulations}
\author[P. Sybilski et al.]
{P. Sybilski,$^{1}$\thanks{E-mail:sybilski@ncac.torun.pl}
M. Konacki,$^{1,2}$ S. Koz{\l}owski$^{2}$\\
$^{1}$Nicolaus Copernicus Astronomical Center, Polish Academy of Sciences, Rabia\'nska 8, 
     87-100 Toru\'n, Poland\\
$^{2}$Astronomical Observatory, Adam Mickiewicz University, S{\l}oneczna 36, 60-186 Pozna\'n, Poland}

\begin{document}

\date{Accepted ... Received ...; in original form ...}

\pagerange{\pageref{firstpage}--\pageref{lastpage}} \pubyear{2010}

\maketitle

\label{firstpage}

\begin{abstract}
The presence of a body in an orbit around a close eclipsing binary star manifests itself through the 
light time effect influencing the observed times of eclipses as the close binary and the circumbinary 
companion both move around the common centre of mass. This fact combined with the periodicity with 
which the eclipses occur can be used to detect the companion. Given a sufficient precision of the 
times of eclipses, the eclipse timing can be employed to detect substellar or even planetary mass 
companions.

The main goal of the paper is to investigate the potential of the photometry based eclipse timing 
of binary stars as a method of detecting circumbinary planets. 
In the models we assume that the companion orbits a binary star
in a circular Keplerian orbit. We analyze both
the space and ground based photometry cases. In particular, we study the usefulness of the on-going 
COROT and Kepler missions in detecting circumbinary planets. We also explore
the relations binding the planet discovery space with the physical parameters of the binaries 
and the geometrical parameters of their light curves. 
We carry out detailed numerical simulations of the eclipse timing by employing a
relatively realistic model of the light curves of eclipsing binary stars. We
study the influence of the white and red photometric noises on the timing
precision. 
We determine the sensitivity of the eclipse timing technique to circumbinary planets for the ground 
and space based photometric observations. We provide suggestions for the best targets, observing 
strategies and instruments for the eclipse timing method. Finally, we compare the eclipse timing 
as a planet detection method with the radial velocities and astrometry.
\end{abstract}


\section{Introduction}
\label{sec1}
Accurate light curves of eclipsing binary stars can be used to precisely measure
the times of eclipses. Such eclipse timing measurements (ET) can then be compared
with the predicted ones and used to infer information on e.g. the presence of an
additional body orbiting the eclipsing binary. The presence of an additional body    
will cause the motion of the eclipsing binary with respect to the centre of mass of
the entire system and result in advances/delays in the times of eclipses due to 
the light time effect. This old idea (it dates back to XVII century and Ole    
Roemer) has been used to e.g. detect stellar companions to eclipsing binaries. 
It can also be used to detect circumbinary planets (P-type planets,
\citealt{dvorak}). Clearly, this idea is simple and has already been explored in the 
literature as a potential way of detecting extrasolar planets 
\citep[see e.g.][]{muterspaugh07a,doyle04a,deeg08a}. 

In this paper, we carry out detailed numerical simulations of ET to explore in 
more depth what can be achieved with this technique both from the ground and space.
In the first part, we study the CoRoT and Kepler and investigate how their
very high photometric precision \citep[][]{alonso,koch04b}
can be used to detect circumbinary planets via ET. In the second part, we
estimate the influence of the red noise and the gaps in the light curves 
typical for the ground based photometry caused by e.g. the day-night cycle, 
technical problems and weather conditions on the discovery space. 

In section \ref{sec2} we describe the light curve and noise models used in the simulations,
in section \ref{sec3} we describe how a planetary timing signal is generated and
detected in a simulated light curve, in section \ref{sec4} we analyze the
space missions CoRoT and Kepler, in section \ref{sec5} we discuss a ground
based effort and conclusions are provided in section \ref{sec6}.

\section{Light curve of an eclipsing binary and its noise}
\label{sec2}
\begin{figure}
\includegraphics[scale=0.25]{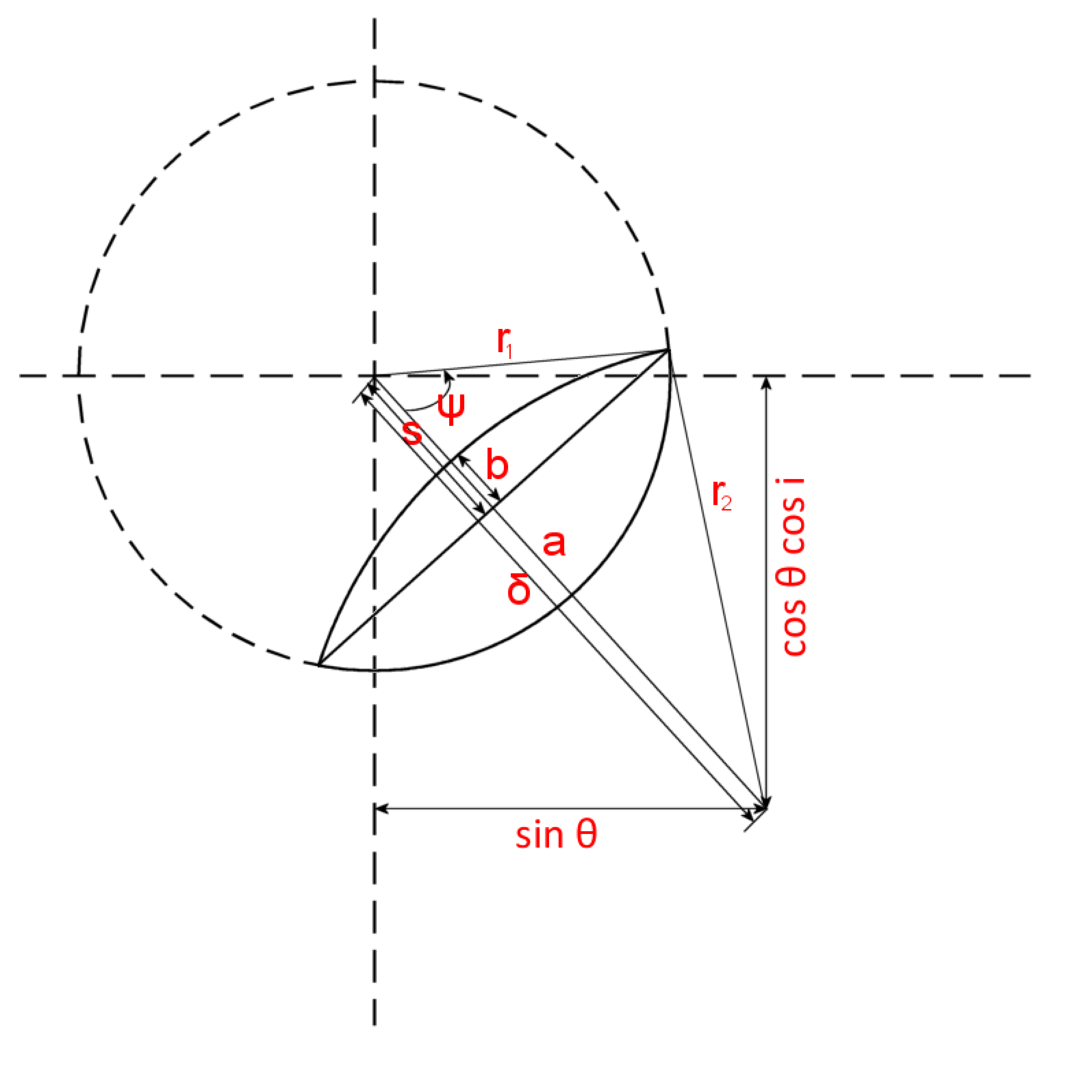}
\caption{The first type of an occultation when the projected stars' separation is 
larger than the radius of the bigger star. The symbols are described in the text 
and in \citealt{nelson}.}
\label{fig:01_geometry1}
\end{figure}

\begin{figure}
\includegraphics[scale=0.25]{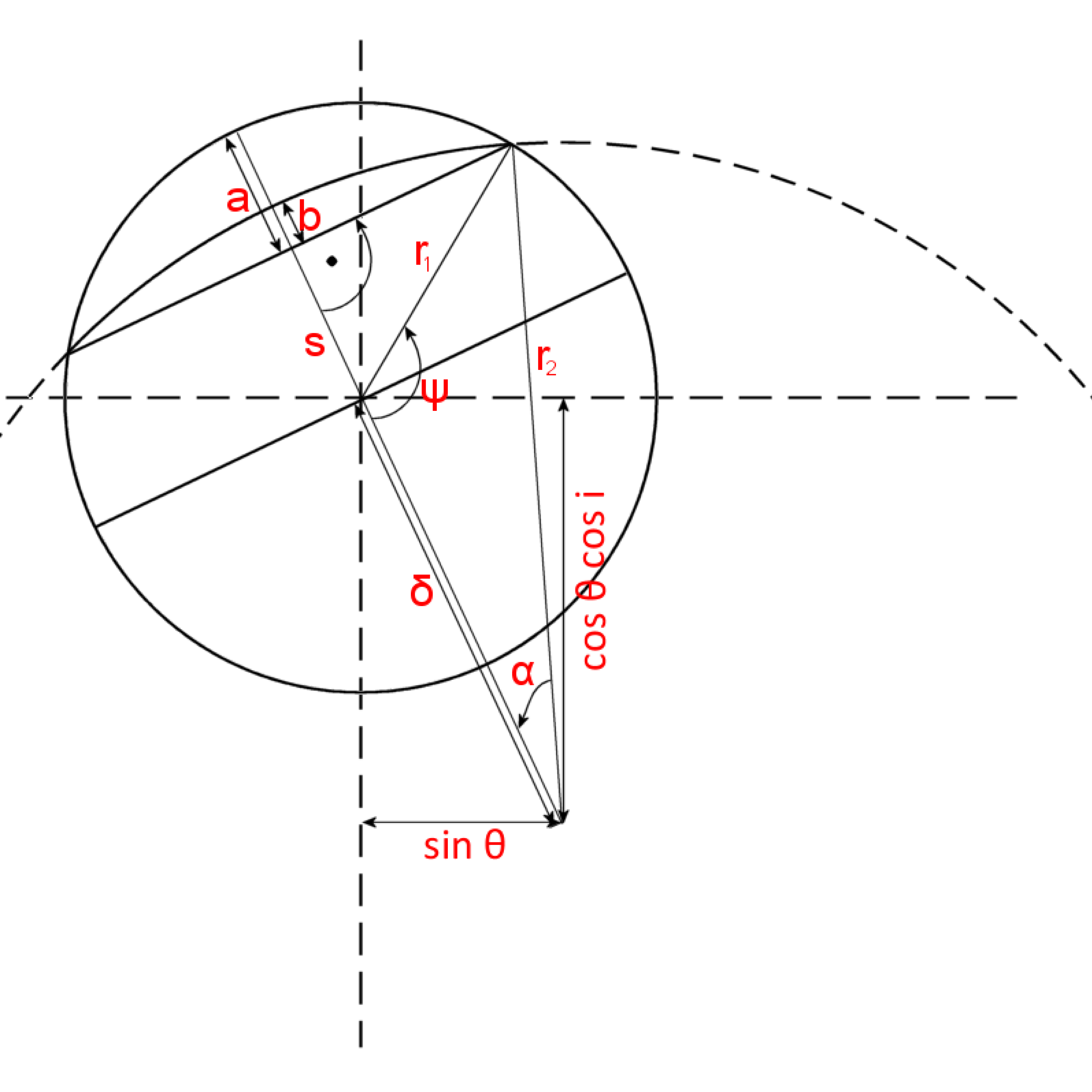}
\caption{The second type of an occultation, the projected stars' separation is smaller than 
the radius of the bigger star.}
\label{fig:02_geometry2}
\end{figure}

The model of an eclipsing binary is from \cite{nelson}. It describes systems that 
do not fill their Roche lobes. We decided to examine detached binaries as
such systems are the most likely ones to serve as stable clocks.  
The adopted model is simple enough to provide fast computations 
and at the same time enables an adequate description of an eclipsing binary. 
Eclipses are described as an obscuration of two discs. The algorithm used to compute 
a synthetic light curve comes from \cite{nelson} and is
based on a few simple equations. Let us note that two of the equations are misprinted 
in \cite{nelson}. The correct version for the eclipsed surface is
\begin{eqnarray}
B = r_1^2\arccos{\frac{r_1-a}{r_1}}-(r_1-a)(2r_1 a-a^2)^{1/2} \\
\qquad+r_2^2\arccos{\frac{r_2-b}{r_2}}-(r_2-b)(2r_2 b-b^2)^{1/2} \nonumber
\end{eqnarray}
for the situation shown in Fig.~\ref{fig:01_geometry1} when the projected
stars' separation is larger than radius of the bigger star and
\begin{eqnarray}
B = \pi r_1^2-r_1^2\arccos{\frac{r_1-a}{r_1}}+(r_1-a)(2r_1a-a^2)^{1/2} \\
+r_2^2\arccos{\frac{r_2-b1}{r_2}}-(2r_2 b - b^2)^{1/2} \nonumber
\end{eqnarray}
for the situation shown in Fig.~\ref{fig:02_geometry2}. In the above $B$ is the obscured 
surface, $r_1$, $r_2$ denote the radius of the first and second star, $a$, $b$ are 
shown in Fig. \ref{fig:02_geometry2} and $\cos\alpha=\frac{r_2-b}{r_2}$, 
$\cos\Psi=\frac{r_1-a}{r_1}$. The remaining symbols are as in the original article.


In the above description, the local perspective effect is ignored. In a typical realistic case 
the separation between the stars is very small compared to the distance between the observer and 
binary. The parameters describing the system are the separation $a$, radius of
the first and second star $r_1$, $r_2$, the total luminosity $L$,  the
fraction of light emitted by second component $L_2$, the Keplerian elements of
the planetary and binary orbits (ellipticity, inclination, semi major axis). Both
orbits, of the planet and the binary star are Keplerian. We assume that the perturbing
planet changes only the position of the binary stars' centre of mass and none of the orbital 
elements. In our simulations we assume circular orbits.

In general we add three types of noise to the synthetic light curves. The first one is 
the photon noise depending on the brightness of an observed eclipsing binary, the second 
one is the white noise of different origin than the photon noise and the third one is the 
red noise due to e.g. the Earth's atmosphere, the instrumentation. The first two are 
typical for the space based photometry and all three are expected to be present in the 
ground based data. 

\begin{figure}
\includegraphics[width=\columnwidth]{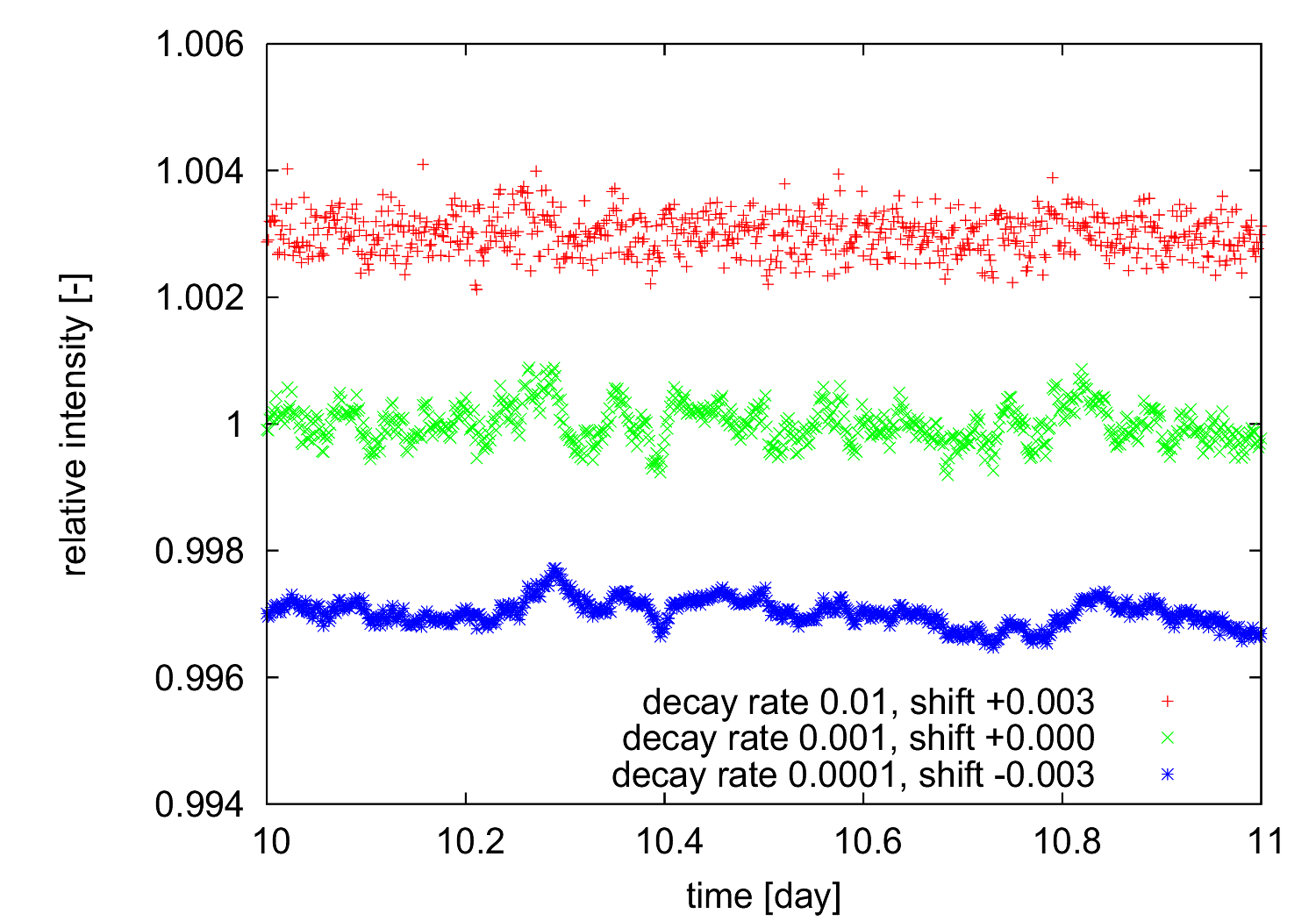}
\caption{Three examples of the red noise with the same standard deviation and different 
decay rates $\lambda_{\mathrm{min}}$ of 0.01, 0.001 and 0.0001. Red noise used in
our simulations of ground observations is characterized by $\lambda_{\mathrm{min}}=$~0.01.}
\label{fig:03_redNoise1}
\end{figure}

The red noise is applied using the package for the exact numerical simulation of power-law 
noises PLNoise \citep[][]{milotti, milotti07}. It is worth noting that 
the impact of the red noise on the planet discovery space differs with 
the typical time scale associated with the red noise. In our code the red noise is parameterized 
via the minimal decay rate of the oscillators generating the noise. Even if the standard deviation 
remains the same, the eclipses timing precision changes with the typical time scale of
the red noise. The examples of red noise with different time scales are shown in 
Figure \ref{fig:03_redNoise1}. 

\section{Planet discovery space}
\label{sec3}

We simulate a light curve of an eclipsing binary with the photon noise
dependent on the brightness of a target and additional Gaussian and red
noises if necessary. The level of the last two is based on the instrument's 
characteristic and the place of the observations (space/ground).
For such a light curve, a planetary light time disturbance affecting the times
of photometric measurements is also added according to the assumed
orbital parameters of a circumbinary planet.

The criterion for a detection or a non detection of a planet is as follows
\begin{equation}
w_j=\frac{1}{\sigma_{x,j}^2}, \quad
\overline{x}_w=\frac{\sum_{j=1}^{L}w_j x_j}{\sum_{j=1}^{L}w_j}, \\
\end{equation}
\begin{equation}
\label{eq06:DetCrit2}
\sigma_S=\sqrt{\frac{\sum_{j=1}^{L}(\overline{x}_w-x_j)^2}{L-1}}\geq 3 max(\sigma_{x,1}, 
\dots , \sigma_{x,L})
\end{equation}
where $x_j$ is the measured time of the $j$-th eclipse (the $t_0$ moment of one 
complete light curve), $\sigma_{x,j}$ denotes its 
standard deviation and $w_j$ the corresponding weight of such a timing measurement. 
Finally $\sigma_S$ is calculated to estimate the magnitude of the timing 
signal which is present in data and this is compared to the largest error of 
$x$. Such a procedure allows us to get a quick insight whether a planetary 
timing signature may be present in the data set. If the inequality
\ref{eq06:DetCrit2} is satisfied it means that a planet is detected
and this is denoted with red colour. Otherwise, a non detection is denoted
with blue colour. The resulting discovery space is a result of an averaging
over a rectangle of 121 neighbouring points. Pure red colour denotes
a certain detection (fraction 121/121) and pure blue colour a certain
non-detection. Intermediate colours are computed as a linear
interpolation between the two basic colours according to a given
fraction of detections in a rectangle (see Figure~4).

Note that such a definition of the detection means that a planetary signal is 
detected when its timing amplitude $A$ is equal to or larger than $\sim4\sigma$ 
where $\sigma$ is the precision (formal error) of the timing accuracy. Hence, 
the main factor that defines the discovery space is the precision with which 
one can measure the moment of one eclipse. 
Obviously, the longer the data set of timing measurements, the
higher is the confidence level with which a planetary signal
is detected \citep[for details see][]{Cum:04::}. For example, since 
Kepler will provide $\sim$10 times longer data sets than CoRoT,
the confidence level of its putative detection would be higher.
Or in other words, for the same confidence level Kepler would
allow for a detection of smaller amplitudes than CoRoT. We have decided to use
our simpler, more conservative approach which is not affected by a particular 
choice of sampling of the timing measurements.

In the above, the time of an eclipse is computed in two ways. The first and classic
approach is to use $x_j$ as one of the parameters of a multiparameter least-squares 
fit of a physical model to the synthetic light curve. 
In this approach the parameters of the binary assumed to calculate the light curve 
are disturbed and used as the starting values of the least-squares 
parameters. Parameters which are varied and then fitted for include the
radii $r_1$, $r_2$, the inclination $i$, $t_0$ the orbital reference epoch,
the orbital period of the binary $P$ and the luminosity of the secondary $L_s$.
We use MINPACK library \citep[][]{more, more84a}\footnote{www.netlib.org}
to carry out a least-squares fitting. 

\begin{figure*}
\centering
\subfigure[]{
\includegraphics[width=\columnwidth]{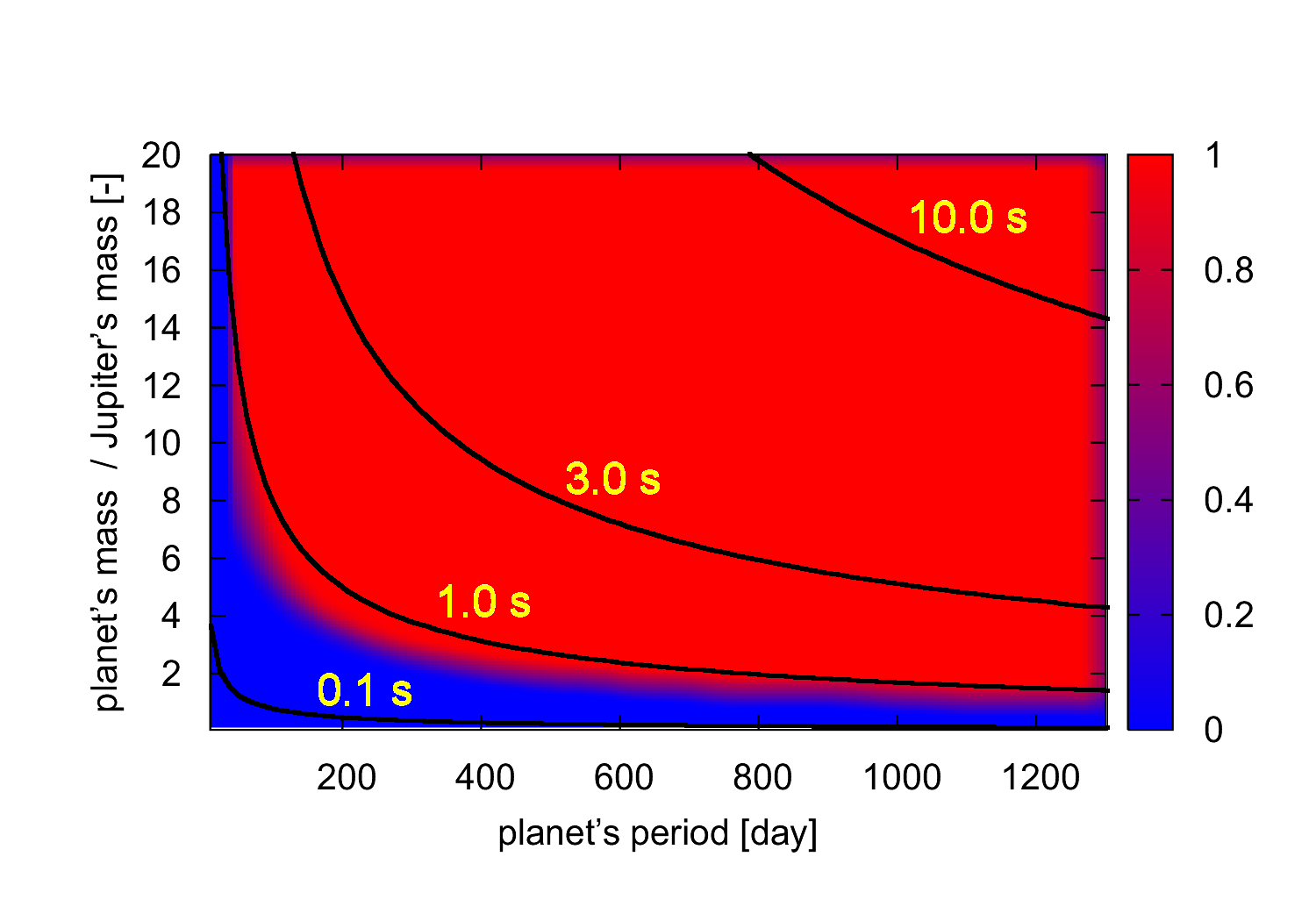}
}
\subfigure[]{
\includegraphics[width=\columnwidth]{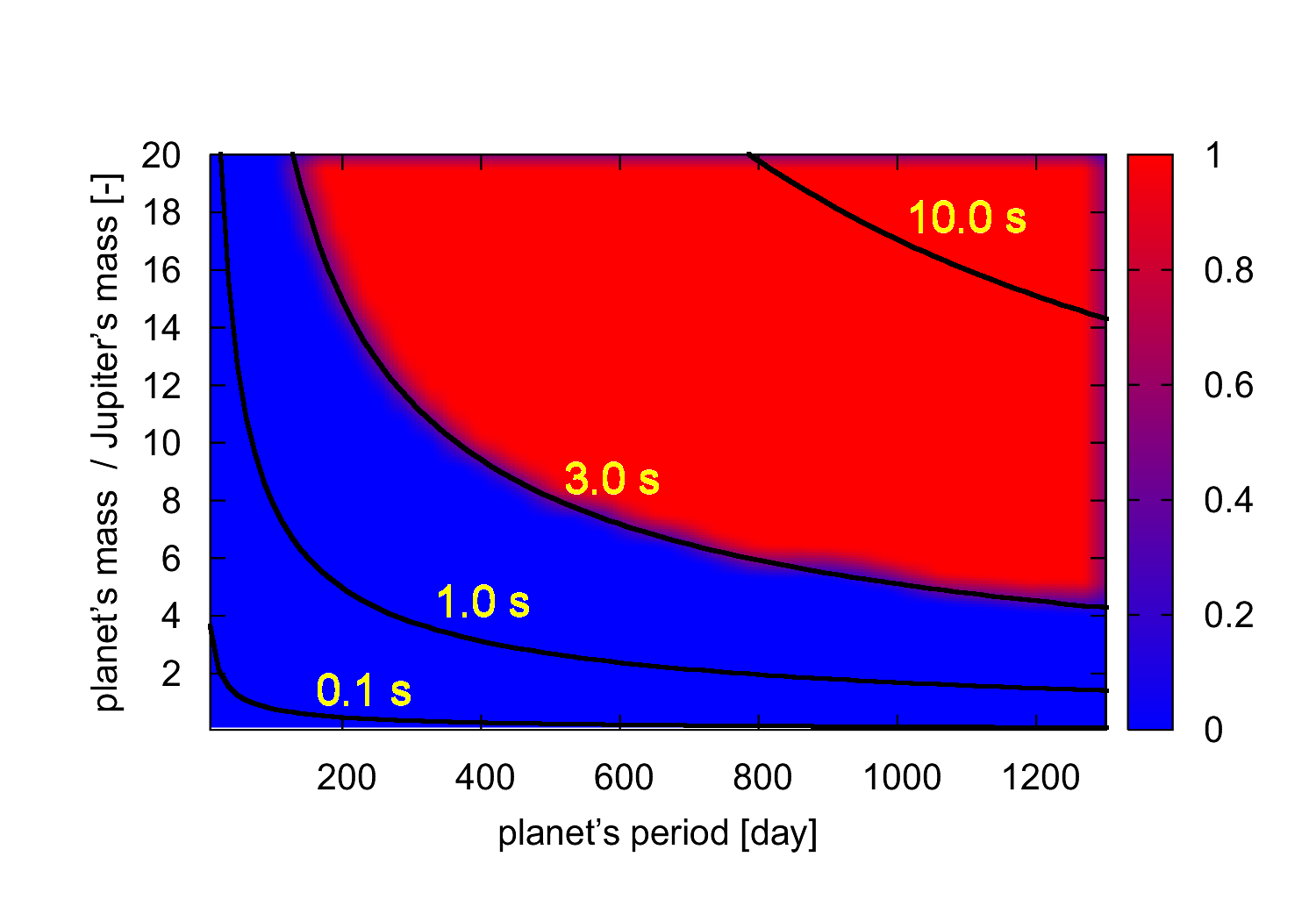}
}
\subfigure[]{
\includegraphics[width=\columnwidth]{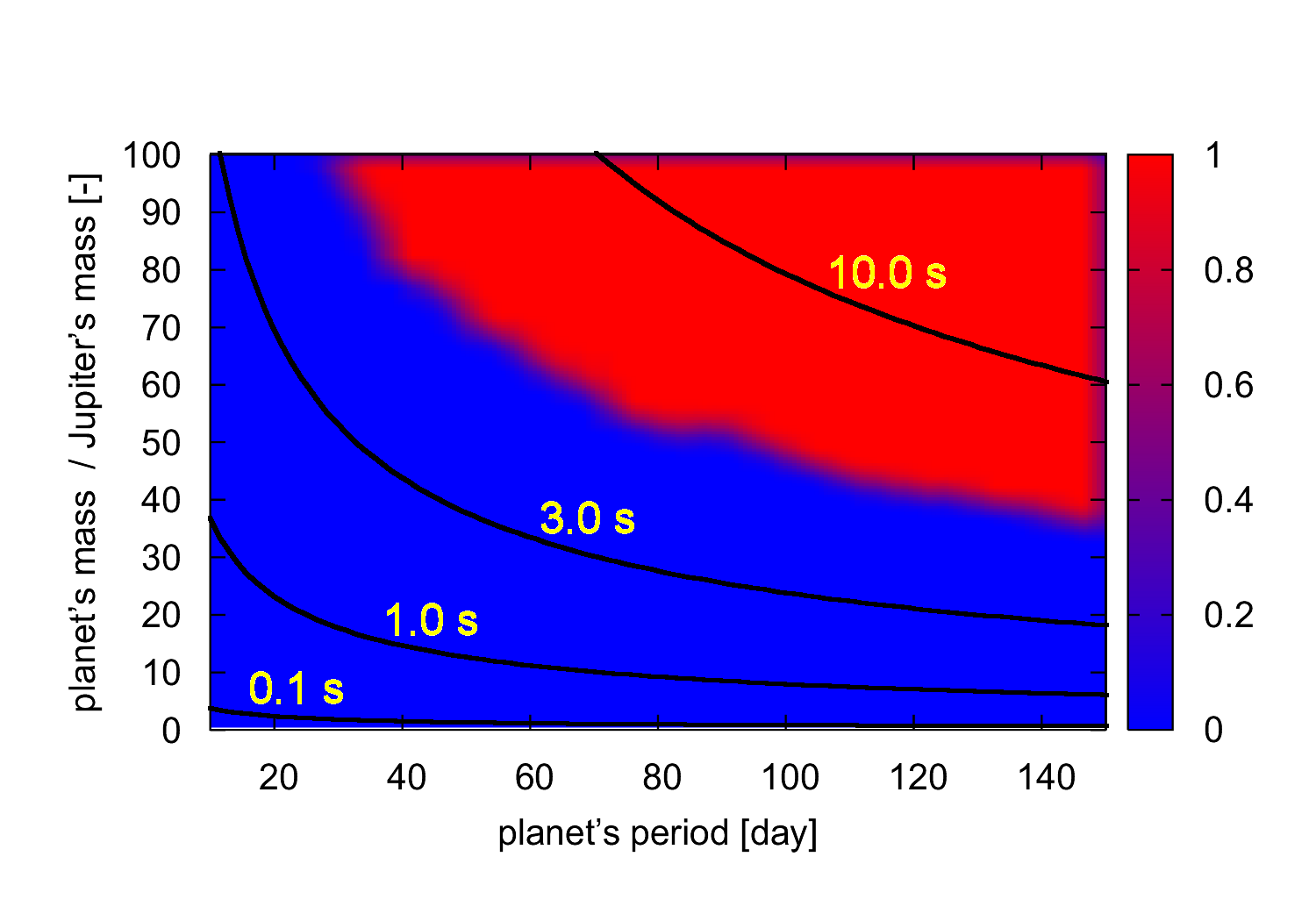}
}
\subfigure[]{
\includegraphics[width=\columnwidth]{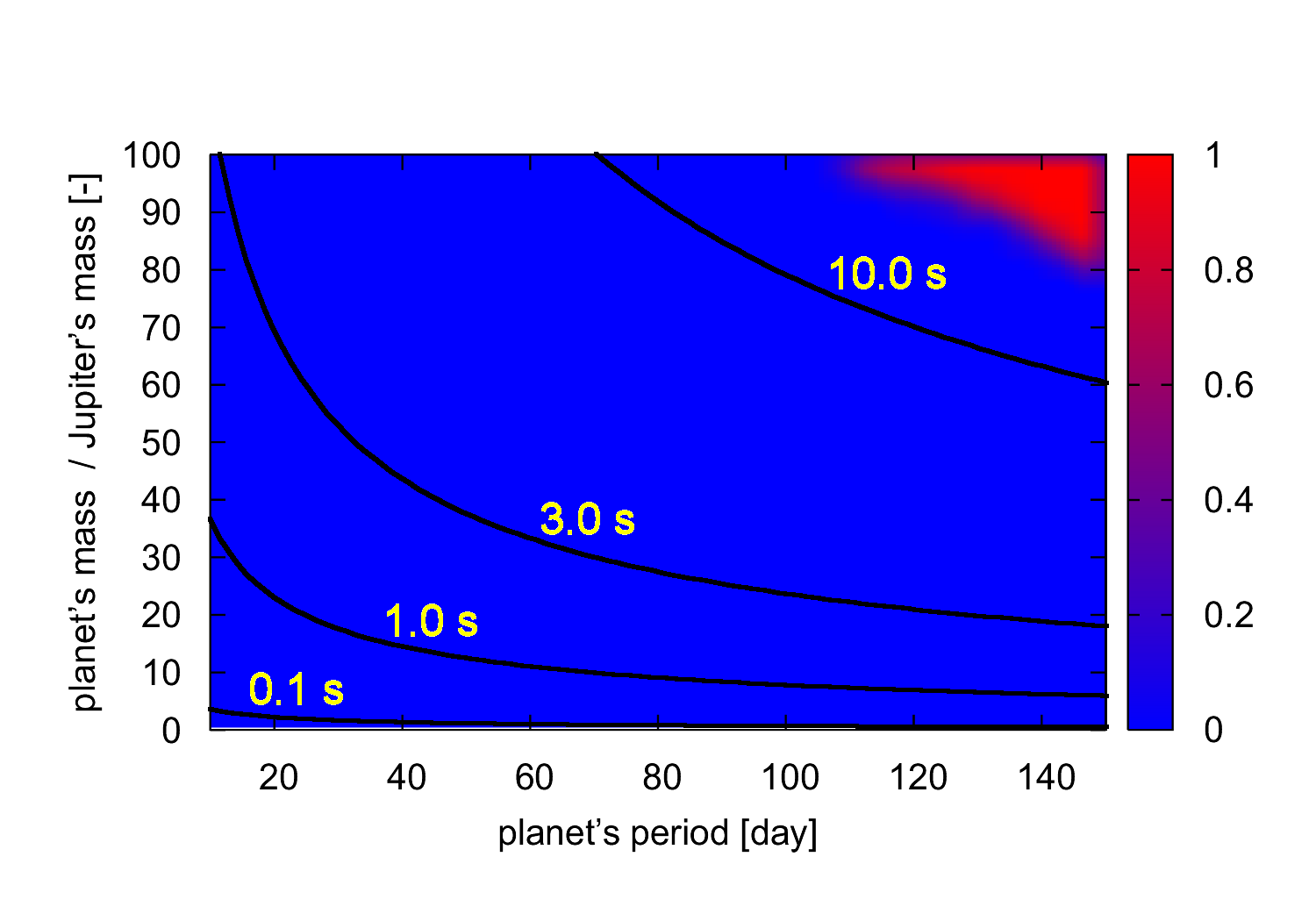}
}
\caption{Typical discovery space for Kepler (top) and CoRoT (bottom). 
Top left for a 9 mag target (the photometric error is $\sigma=0.035$ mmag)
and top right for a 14 mag target ($\sigma=0.17$ mmag).
Bottom left for a 13 mag target ($\sigma=0.5$ mmag) and bottom right
for a 15 mag target ($\sigma=1.5$ mmag). One should remember
that a measurement is taken every 900 sec for Kepler 
which is in the middle of 30 minutes and 1 minute cadence now used
and 320 sec for CoRoT hence the total number of measurements varies per one full orbital
period of the target binary. Now Kepler uses 30 and 1 minute cadences
but this does not affect the outcome of the simulations. For example
longer integration times provide a smaller photometric error but also
a smaller number of measurements per light curve.}
\label{fig:CORKEP}
\end{figure*}

The simulations can be easily extended to e.g. cases with the elliptically distorted 
stars and include the limb darkening as e.g. in \cite{nelson}. 
As we have tested, in such a case the results are slightly different after introducing 
these two effects and the total time required to compute a discovery space is a few 
times longer. The resulting area of the discovery space corresponding to detectable 
planets is a bit smaller and moves toward upper-right corner 
(see e.g. Figure \ref{fig:CORKEP}). This is in our
opinion the result of the correlations between the increasing number of
parameters used in the least-squares fit which often accounts for very subtle
effects. For this reason, we believe that it is best to use an approach
known from e.g. radio pulsar timing and precision radial velocities
relying on a reference template pulse or spectrum to measure a timing or
RV shift.

In such an approach the time of an eclipse is computed by comparing a given light curve 
with a reference light curve obtained by folding all the simulated photometric 
measurements with the orbital period of a binary. As mentioned, this is an approach 
used in radio pulsar timing or in precision radial velocity technique where the 
cross correlation function and the reference radio pulse or template spectrum are 
used to compute a timing shift or a Doppler shift. In our case, in order to 
compute $x_j$ and its formal error  $\sigma_{x,j}$ we use the least-squares 
formalism. Let us note that for the simpler light curve model with the spherical
binary components and no limb darkening, both approaches result
in the same discovery space and for the more complicated binary model case, 
the approach employing a reference light curve results in a better
discovery space (a wider range of detectable planets). Obviously the second
approach will work well only if the light curve is sufficiently stable
but then only for such stable light curves/binaries one may hope to detect 
planets.

The parameters of the binary star, the instruments and the simulated
photometric measurements are summarized in Tables \ref{tab:10_starPar}, 
\ref{tab:09_SimInput} and \ref{tab:11_SimPar}.
The discovery space is computed on a dense grid of planetary periods and
masses. For the figures, the discovery space is represented by averaging 121
neighbour points from the grid. The red colour in the diagrams denotes certain 
detection (fraction 121/121) and the blue one the lack of detection (0/121). 
All the intermediate colours are computed as a linear interpolation. Black lines 
in the discovery space show the planet's mass and period which generates 
a given timing amplitude, $A$. The equation describing such a line is given by
\begin{equation}
\label{eq05:DSA}
M_P(P_{pl}) = \left( \frac{4 \pi^2 M_{B}^2}{P_{pl}^2 G} \right)^{\frac{1}{3}}\cdot(Ac)
\label{eq.MP}
\end{equation}
where $M_P$, $P_{pl}$ are the mass and period of a planet, $M_{B}$ is the mass of the 
binary star, $c$ stands for the speed of light and $G$ is the gravitational constant. 
\begin{table}
\caption{Binary's star characteristic}
\label{tab:10_starPar}
\centering
\begin{tabular}{r l}
\hline
Total luminosity & 2 L$_{\sun}$ \\
Secondary star luminosity & 1 L$_{\sun}$ \\
Total binary star mass & 2 M$_{\sun}$ \\
Effective temperature of binary components & 5780 K \\
Orbital eccentricity & 0 \\
Radii & 1 R$_{\sun}$ \\
Orbital period & 3 days \\
Inclination & 90 deg \\
Orbit inclination of a planet & 90 deg \\
\hline
\end{tabular}
\end{table}

\begin{table}
\caption{Instrument characteristic used in our simulations.}
\label{tab:09_SimInput}
\centering
\begin{tabular}{r  c  c  c  l}
\hline
Parameter & CoRoT & Kepler & ground & Unit \\
\hline \hline
White noise & 0.07 & 0.02 & 0.35 & mmag \\
Red noise & 0 & 0 & 0.35 & mmag \\
Band & 370-950 & 430-890 & 502-587 & nm \\
Integration time & 320 & 900 & varies & s \\
Observing window & 150 & 1461 & 365 & days \\
Aperture & 588 & 2256 & 1963.5 & cm$^2$ \\
Throughput & 81 & 81 & 81 & \% \\
Target stars (transits) & 12-15.5 & 9-14 & - & mag \\
Target stars & 6-9 & - & - & mag \\
(stellar seismology) & & & & \\
Target stars & - & - & 6-14 & mag \\
(ET) & & & & \\
\hline
\end{tabular}
\end{table}

\begin{table}
\caption{Simulation's internal parameters}
\label{tab:11_SimPar}
\centering
\begin{tabular}{r l}
\hline
Orbital period & from 10 days \\
Planet mass & from 0.05 M$_{\mathrm{Jupiter}}$ \\
Effective no. of simulation points & 40401 \\
No. of light curve active parts & varies \\
Red noise $dt$ & 1 \\
Red noise $nt$ & 0.1 \\
Red noise $\lambda_{\mathrm{min}}$ & 0.01 \\
Red noise $\lambda_{\mathrm{max}}$ & 1 \\
\hline
\end{tabular}
\end{table}

Figure \ref{fig:CORKEP} shows a few examples of a typical discovery space
for the CoRoT and Kepler missions. Each discovery space is a result of one 
simulation run covering 40401 points (201 by 201) for which the inequality 
Eq. \ref{eq06:DetCrit2} was checked. 
Based on a discovery space and Equation \ref{eq.MP}, we derive a timing
amplitude which fits best to the border between a detection and non detection. 
As can be seen e.g. in Figure \ref{fig:CORKEP}(a-b) $A$ is an approximation
of the actual border and hence the best fit value of $A$ is also 
characterized by its non zero error $\sigma_{A}$ depending on how the 
real border deviates from $A$.

\section{CoRoT and Kepler}
\label{sec4}
The ongoing space missions CoRoT and Kepler aimed to detect transiting
planets can in principle be used also to detect circumbinary planets
via ET. These missions have the obvious advantage over any ground effort
of providing an uninterrupted set of photometric measurements of a respectively
150 day and four year time span. In our simulations we used the parameters
of the missions as described by \cite{costes04c}, \cite{garrido06} 
and \cite{auvergne} for CoRoT and by \cite{koch04b} for Kepler.
They are also summarized in Table \ref{tab:09_SimInput}.

\begin{figure}
\includegraphics[width=\columnwidth]{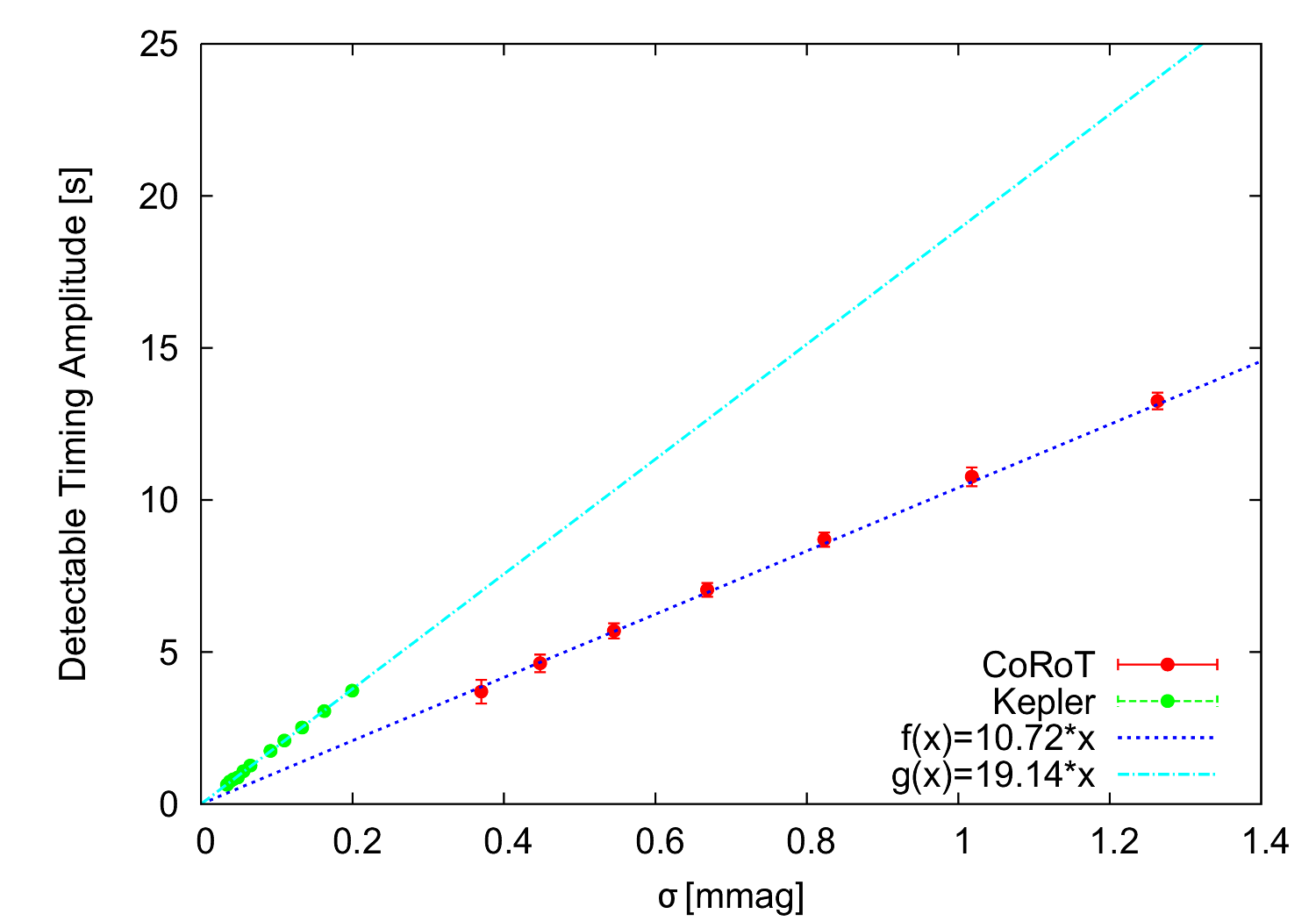}
\caption{DTA for a given photometric precision for CoRoT and Kepler 
missions along with the best fitting relation. Note that the
difference between the two cases is dependent on the cadence
of photometric measuremenets (CoRoT $\sim$320 and Kepler 
$\sim$900 secodns).}
\label{fig:01_sigmaVsA}
\end{figure}

The examples of discovery space for CoRoT and Kepler are shown in 
Figure \ref{fig:CORKEP}. While both instruments are capable of
providing very precise photometry, the resulting discovery space
is also affected by the cadence of photometric measurements
and the brightness of the targets. Overall the potential of
detecting circumbinary planets with CoRoT and Kepler comes out 
somewhat less attractive than one may have hoped for. The simulations
allow us to determine the following relations for CoRoT $A(\sigma)=10.72~\sigma$, 
and for Kepler $A(\sigma)=19.14~\sigma$, $\sigma$ is the photometric
precision of a single measurement in mmag and $A$ the detectable
timing amplitude (DTA) is given in seconds (see Figure \ref{fig:01_sigmaVsA}). 

The potential problem with both missions is the predefined target pool.
If this is taken into account, despite high photometric precision the
chances of detecting a circumbinary, non-transiting planet may be somewhat 
small. For example for the Kepler mission an upper limit to detectable
circumbinary planets may be estimated as follows
\begin{equation}
\label{eq07:howMany}
n_P=n_* p_* p_J = 10^{5} \cdot 0.00016 \cdot 0.06 = 0.96 
\end{equation}
where:
\begin{itemize}
\item $n_*$ is the number of targets, $n_* \approx 10^{5}$ (kepler.nasa.gov),
\item $p_{*}$ is the eclipsing detached binary stars ratio to all stars, 
$p_{*} \approx 0.00016$ 
\citep{paczynski06a},
\item $p_J$ is the percentage of stars having giant planets
(assuming that it is the same as for single stars), $p_J \approx
0.06$
\end{itemize}
and this does not take into account the probability of detecting
a planet with a given mass and period via ET. Clearly, in order
to detect circumbinary planets using ET one may have to turn to the ground
based observations where the target pool can be carefully selected
and fine-tuned to provide for the highest possible chances of 
detecting circumbinary planets. 
However, let us note that in the above we used a percentage
of {\it detached} binaries in the ASAS catalogue. 
If one is willing to try the presumably
less stable contact binaries as well, then $p_{*}$ increases to
$\sim0.00065$ according to the results from the ASAS sample
\citep{paczynski06a}. Even more optimistic preliminary results
come from the Kepler and CoRoT fields for which the percentage
of eclipsing binaries is $\sim0.007$ (H. Deeg, L. Doyle 
personal communication). In such a case, the number of potentially
detectable planets would rise to $\sim40$.

\begin{figure*}
\label{fig:CORKEP2}
\centering
\subfigure[]{
\includegraphics[width=\columnwidth]{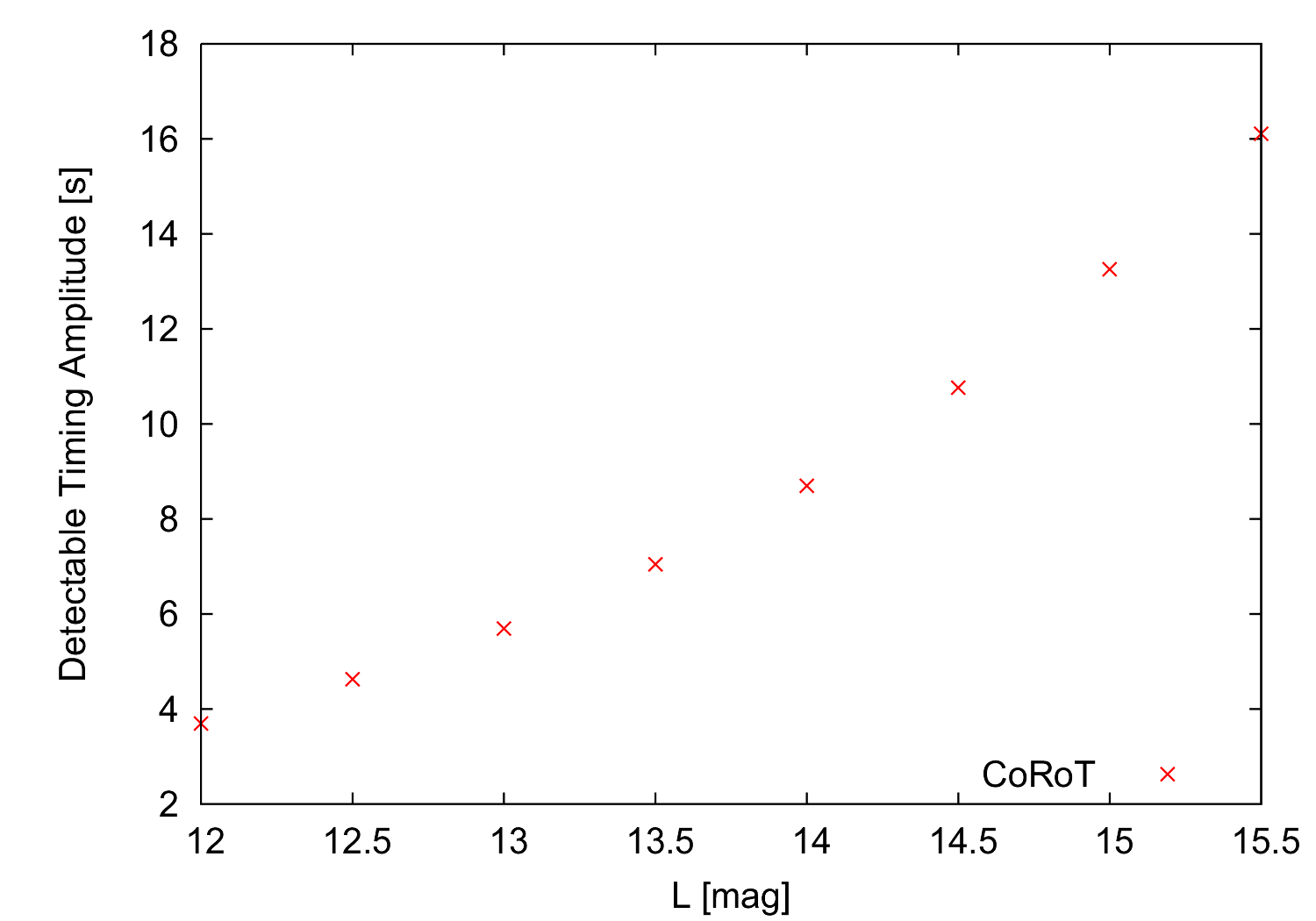}
}
\subfigure[]{
\includegraphics[width=\columnwidth]{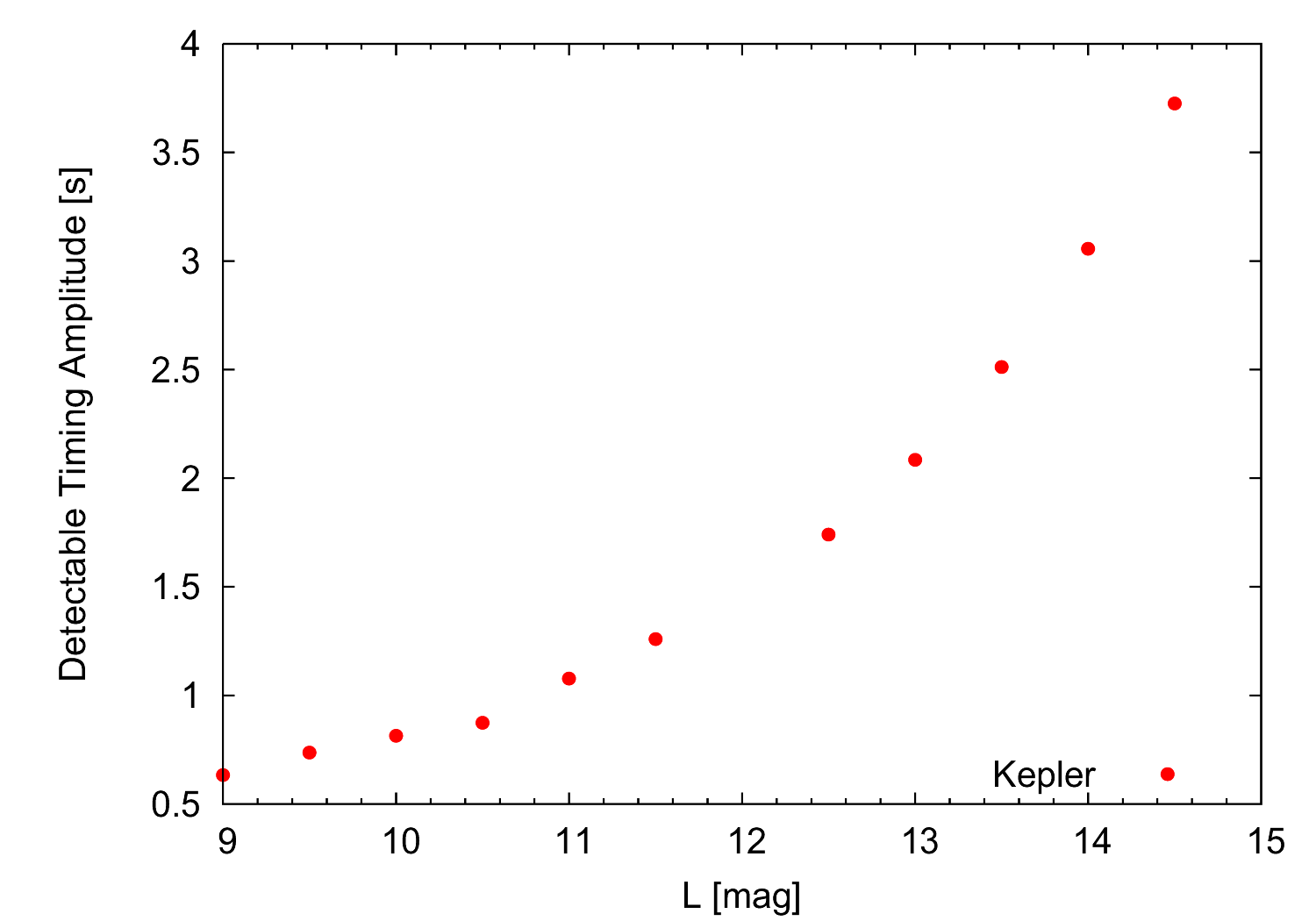}
}
\subfigure[]{
\includegraphics[width=\columnwidth]{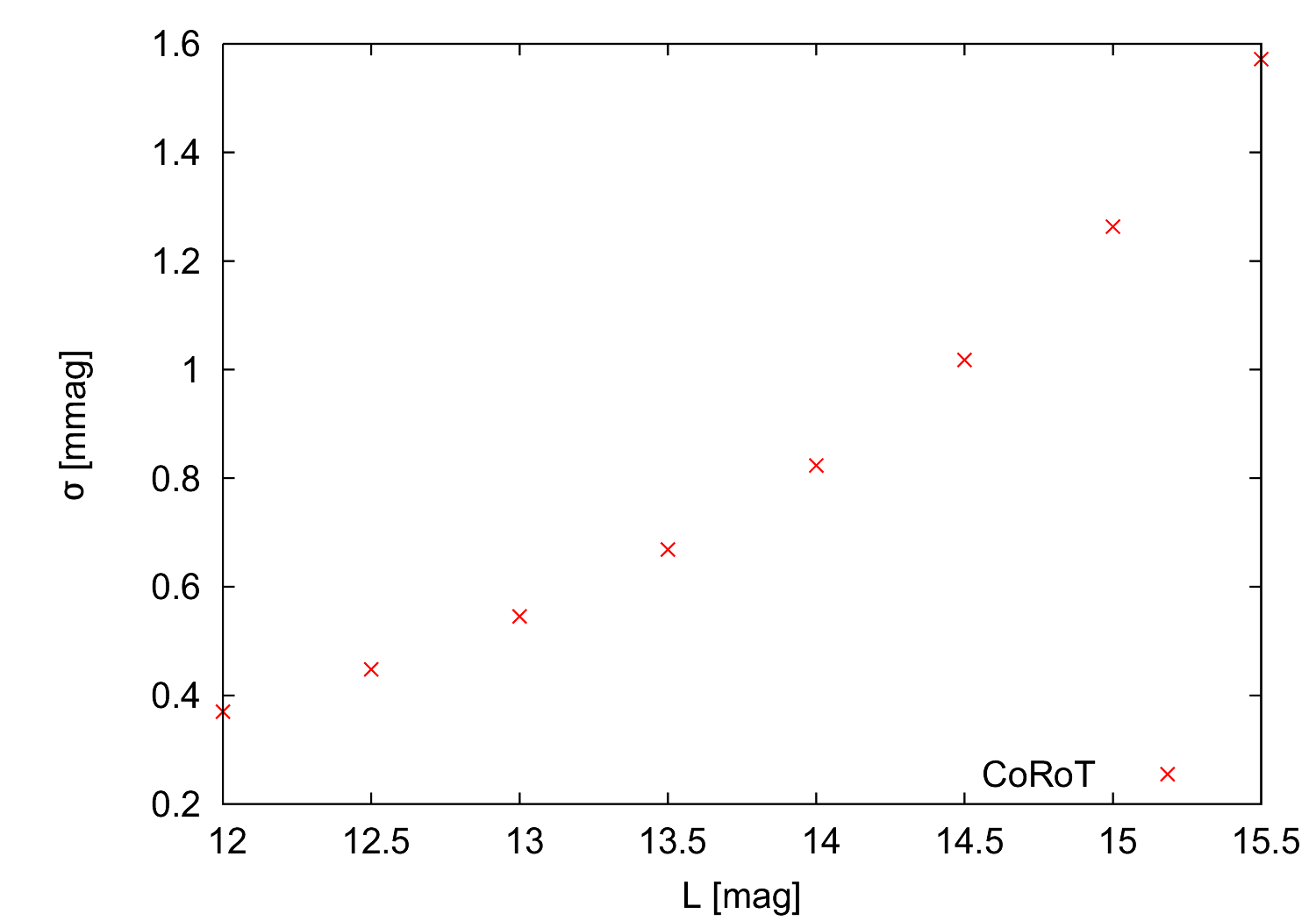}
}
\subfigure[]{
\includegraphics[width=\columnwidth]{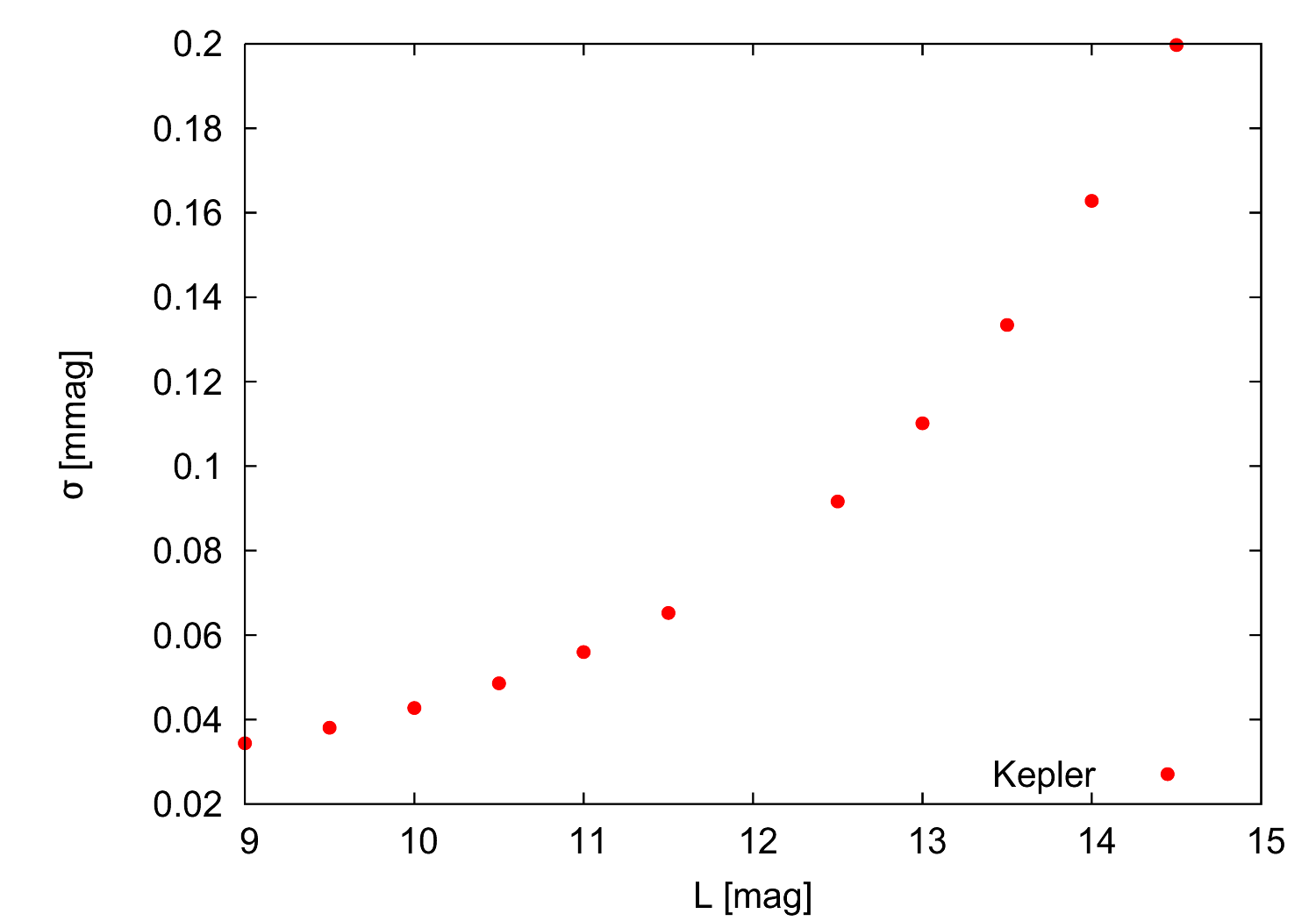}
}
\caption{Typical DTA and photometric error for
CoRoT (left) and Kepler (right) missions and an object
with brightness $L$.}
\end{figure*}

\section{Observations from the ground}
\label{sec5}

Performing numerical simulations that would aim to answer all or
almost all questions related to a search for circumbinary planets
using ground based ET is not practical. Possible observing scenarios
are highly dependent on a choice of a target, its parameters
and the coordinates of an observatory. Such simulations can
be carried out if the target pool and the observatory are already 
chosen. For these reasons, we present below a few typical problems 
one may encounter when performing a ground based ET survey 
and analyze them through numerical simulations. These results
are an essence of a much larger set of simulations we carried out.

\begin{figure*}
\centering
\subfigure[]{
\includegraphics[width=\columnwidth]{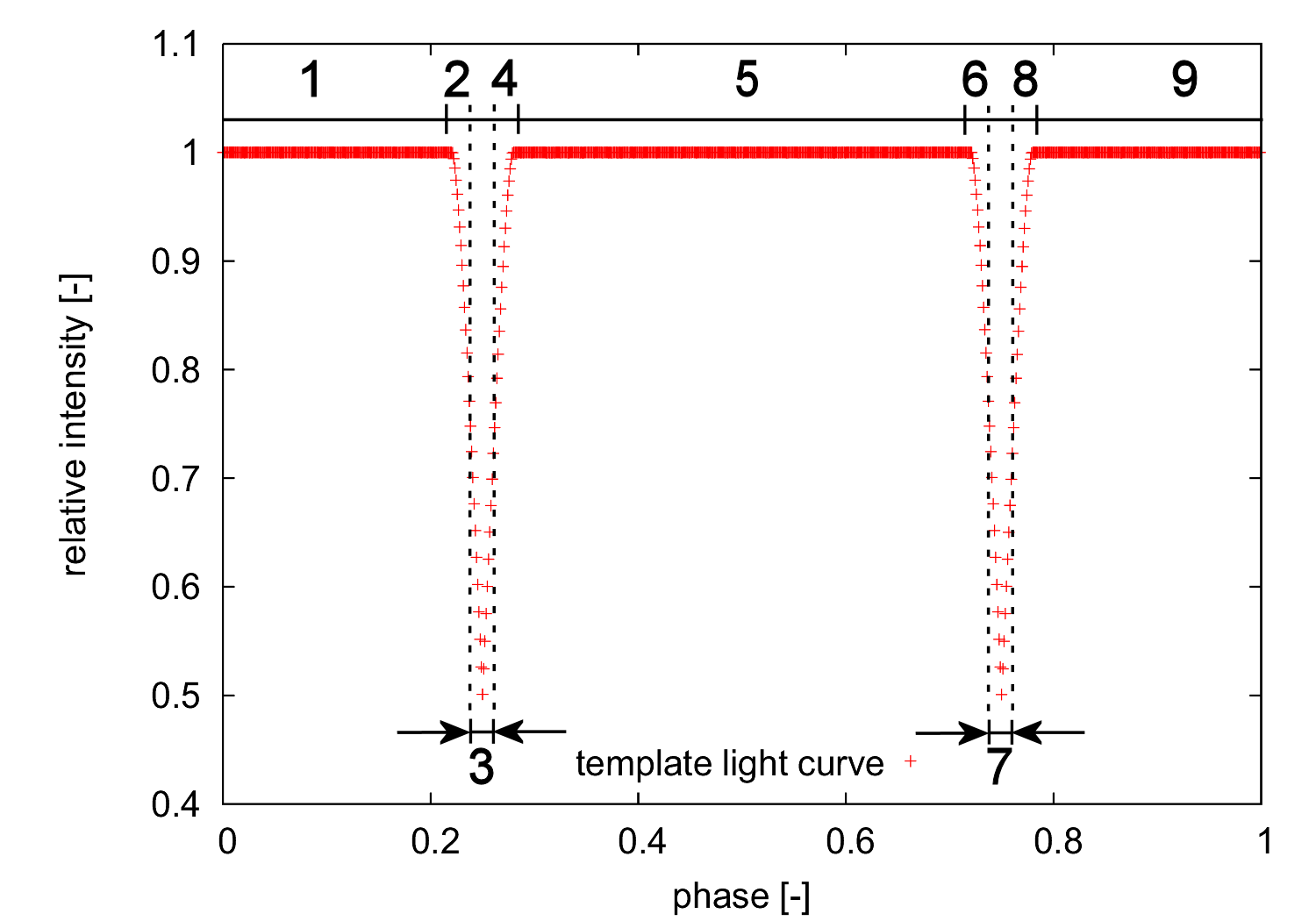}
\label{fig:14_template1}
}
\subfigure[]{
\includegraphics[width=\columnwidth]{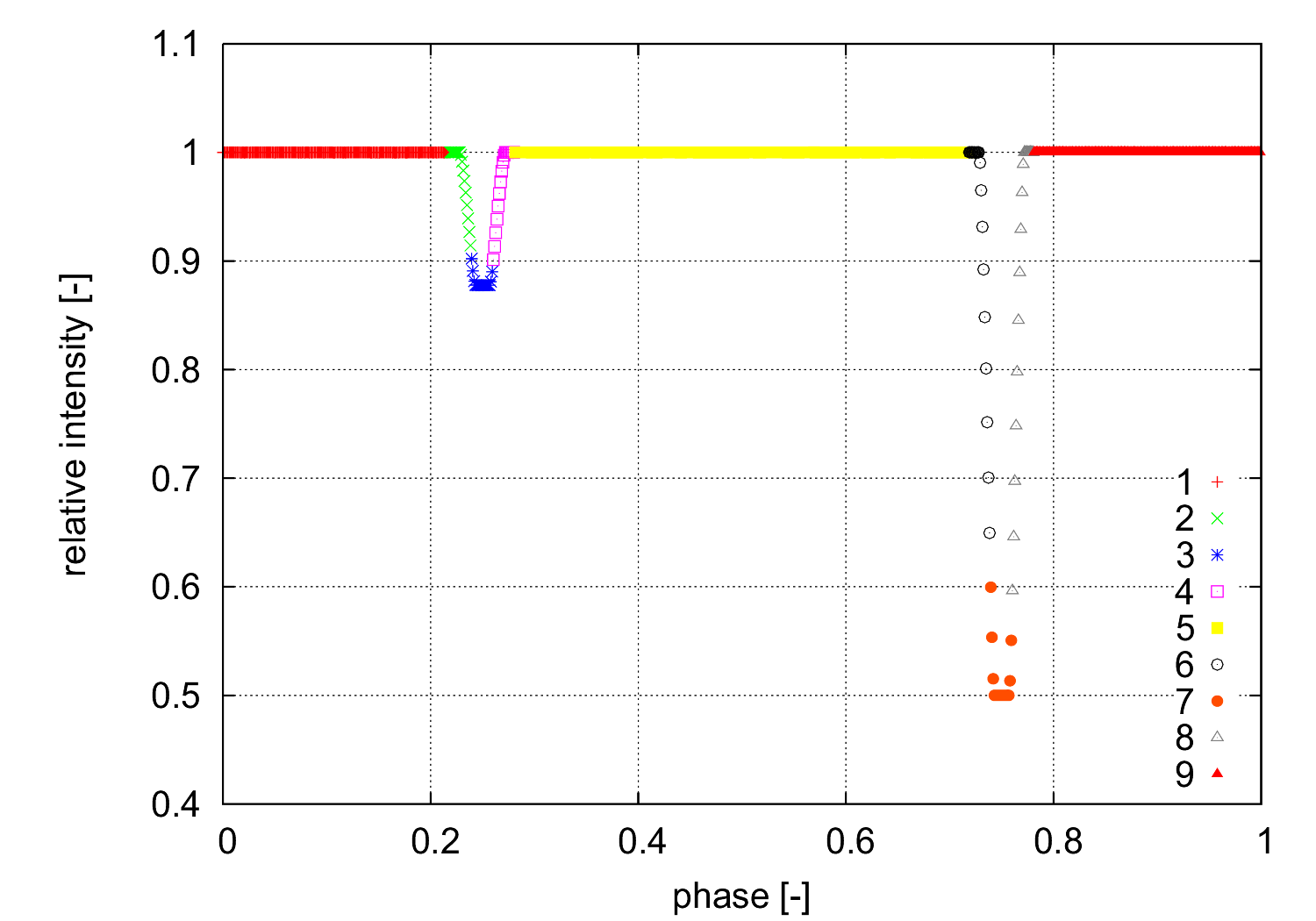}
\label{fig:14_template2}
}
\caption{Light curve templates for simulations exploring an influence of the different 
parts of light curves on DTA. Figure \subref{fig:14_template1} is the template of 
V-shape eclipses, \subref{fig:14_template2} represents flat bottom eclipses.}
\label{fig:14_template}
\end{figure*}

\begin{table*}
\caption{Impact of different parts of a light curve and
their combinations on DTA (in seconds) for a V-shaped
eclipse shown in Figure \ref{fig:14_template1}. 
DTA of 20 sec means that a timing measurement was not possible.}
\label{tab:01_LCelVmean}
\centering
\begin{tabular}{r|*{9}{r}}
no. of active parts \textbackslash~part no. & 1 & 2 & 3 & 4 & 5 & 6 & 7 & 8 & 9 \\
\hline
1	 & 	20.00	 & 	6.28	 & 	3.74	 & 	8.08	 & 	20.00	 & 	8.16	 & 	4.88	 & 	8.36	 & 	20.00	 \\
2	 & 	9.95	 & 	5.27	 & 	3.53	 & 	6.34	 & 	9.96	 & 	6.25	 & 	4.26	 & 	6.54	 & 	9.97	 \\
3	 & 	6.02	 & 	4.50	 & 	3.32	 & 	5.03	 & 	6.01	 & 	4.99	 & 	3.81	 & 	5.13	 & 	6.03	 \\
4	 & 	4.46	 & 	3.91	 & 	3.12	 & 	4.16	 & 	4.45	 & 	4.15	 & 	3.47	 & 	4.20	 & 	4.47	 \\
5	 & 	3.71	 & 	3.47	 & 	2.96	 & 	3.59	 & 	3.71	 & 	3.56	 & 	3.20	 & 	3.59	 & 	3.71	 \\
6	 & 	3.47	 & 	3.36	 & 	3.10	 & 	3.50	 & 	3.58	 & 	3.50	 & 	3.31	 & 	3.52	 & 	3.59	 \\
7	 & 	2.92	 & 	2.87	 & 	2.70	 & 	2.90	 & 	2.92	 & 	2.89	 & 	2.80	 & 	2.90	 & 	2.93	 \\
8	 & 	2.67	 & 	2.65	 & 	2.58	 & 	2.66	 & 	2.67	 & 	2.66	 & 	2.62	 & 	2.66	 & 	2.67	 \\
9	 & 	2.52	 & 	2.52	 & 	2.52	 & 	2.52	 & 	2.52	 & 	2.52	 & 	2.52	 & 	2.52	 & 	2.52	 \\
\end{tabular}
\end{table*}

\begin{table*}
\caption{Impact of different parts of a light curve and their combinations 
on DTA (in seconds) for an eclipse with a flat bottom shown in Figure 
\ref{fig:14_template2}.
DTA of 20 sec means that a timing measurement was not possible.}
\label{tab:03_LCelFmean}
\centering

\begin{tabular}{r|*{9}{r}}
no. of active parts \textbackslash~part no. & 1 & 2 & 3 & 4 & 5 & 6 & 7 & 8 & 9 \\
\hline
1	 & 	20.00	 & 	6.88	 & 	20.00	 & 	20.00	 & 	20.00	 & 	4.29	 & 	9.60	 & 	4.21	 & 	20.00	 \\
2	 & 	12.89	 & 	6.76	 & 	11.33	 & 	11.46	 & 	14.51	 & 	3.80	 & 	8.10	 & 	3.87	 & 	12.89	 \\
3	 & 	8.52	 & 	6.75	 & 	7.61	 & 	7.58	 & 	9.52	 & 	3.52	 & 	5.43	 & 	3.65	 & 	8.51	 \\
4	 & 	5.14	 & 	4.87	 & 	5.14	 & 	4.85	 & 	5.15	 & 	3.29	 & 	4.51	 & 	3.39	 & 	5.17	 \\
5	 & 	4.77	 & 	4.49	 & 	4.56	 & 	4.53	 & 	4.72	 & 	3.16	 & 	3.91	 & 	3.22	 & 	4.72	 \\
6	 & 	3.54	 & 	3.48	 & 	3.51	 & 	3.52	 & 	3.54	 & 	2.93	 & 	3.36	 & 	3.00	 & 	3.53	 \\
7	 & 	3.06	 & 	3.04	 & 	3.05	 & 	3.06	 & 	3.06	 & 	2.78	 & 	3.00	 & 	2.82	 & 	3.06	 \\
8	 & 	2.76	 & 	2.75	 & 	2.75	 & 	2.75	 & 	2.75	 & 	2.65	 & 	2.74	 & 	2.66	 & 	2.76	 \\
9	 & 	2.55	 & 	2.55	 & 	2.55	 & 	2.55	 & 	2.55	 & 	2.55	 & 	2.55	 & 	2.55	 & 	2.55	 \\
\end{tabular}
\end{table*}

The two main differences between the space and ground based surveys are 
the presence of red noise due to predominately the atmosphere and the 
typically very incomplete light curves one obtains with a ground based 
telescope for a detached eclipsing binary that typically has an orbital
period much longer than the duration of a night. For our reference
binary (Table~1), we simulated two types of light curves. One
with V-shaped eclipses (the parameters exactly as in Table~1)
and one with eclipses with flat bottoms. In the later case,
we assumed a radius ratio of 0.5 while keeping $L_2 = 0.5$.
To the simulated photometric measurements we added red noise
with an RMS of 0.35 mmag, white noise with an RMS of 0.35 mmag
and photon noise depended on the actual momentary brightness.
We divided the light curves into 9 parts to explore which 
parts when present have the largest impact on the detectable
timing amplitude (see Figures \ref{fig:14_template1} and 
\ref{fig:14_template2}). In the simulations we removed parts 
of the light curves and used such incomplete
light curves to measure the times of eclipse. 
All the possible permutations without repetition were 
used for sets ranging from 1 to 9 light curve's elements. 
Afterwards an average DTA was calculated.  Whenever a DTA could not 
be computed, the result by default was set to 20 sec.
The results are shown in Tables \ref{tab:01_LCelVmean}, 
and \ref{tab:03_LCelFmean}. In the tables the columns
correspond to the parts of light curve which were for
sure present and the rows correspond to the total
number of parts present ("active") in a light curve. 
This approach allows us to easily determine the most valuable 
parts of a light curve and establish which sets of them 
when present provide even better results. 

In the case of V-shaped eclipses the most important are the parts 3 and 7 in 
Figure \ref{fig:14_template1}
which correspond to the middle parts of eclipses. Next come ingress and 
egress. For the eclipses with flat bottoms the most valuable parts
are 6 and 8 in Figure \ref{fig:14_template1} (ingress and egress 
of the deeper eclipse).  
Altogether most of the timing information can be
derive by just observing the entire eclipses. This is consistent with 
common sense and is doable from the ground. 
Let us also note that in the simulations we used the first type 
of red noise from Figure~3. We also determined that its impact on the detectable
timing amplitude is comparable to white noise with and RMS twice as
large (i.e. 0.7 mmag).

\begin{figure}
\includegraphics[width=\columnwidth]{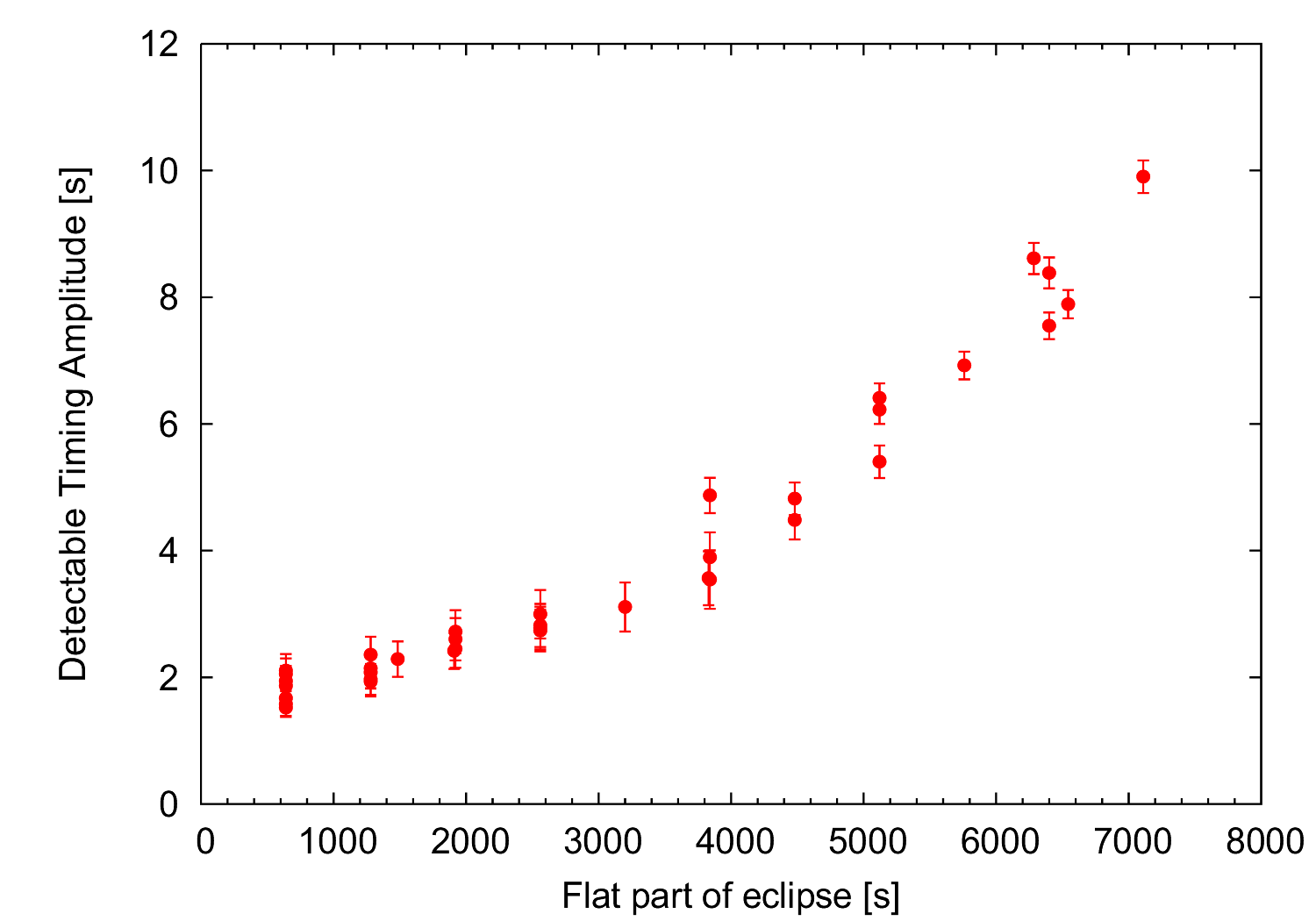}
\caption{Duration of the flat part of an eclipse and 
DTA.}
\label{fig:10_efFlat}
\end{figure}

\begin{figure}
\includegraphics[width=\columnwidth]{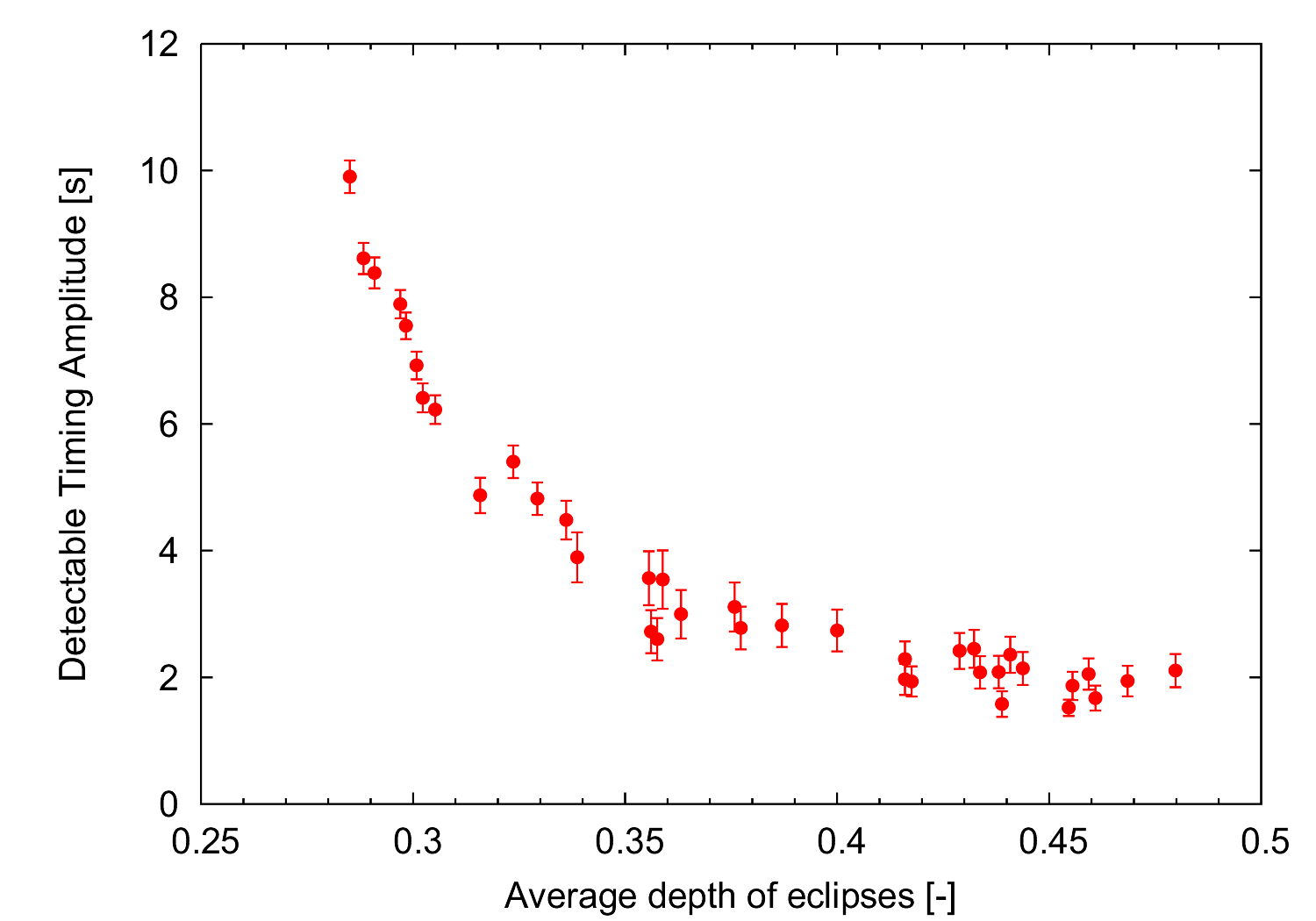}
\caption{Depth of an eclipse and DTA.}
\label{fig:11_efHeight}
\end{figure}

In another set of simulations we tested a number of features of 
an eclipse that affect the timing precision and detectable 
amplitude. We tested the impact of the duration of a flat
part of an eclipse (Figure~\ref{fig:10_efFlat}), the impact
of the eclipse's depth (Figure~\ref{fig:11_efHeight}),
the impact of the duration of a V-shaped eclipse, the impact of
the secondary star's luminosity (Figure~\ref{fig:09_efLs}) and the radius
ratio (Figure~\ref{fig:04_RadiusDSE}) on the detectable timing amplitude (DTA).
The conclusion is that it is preferable to observe
V-shaped short lasting and deep eclipses to maximize
DTA which is not an unexpected result.

\begin{figure}
\includegraphics[width=\columnwidth]{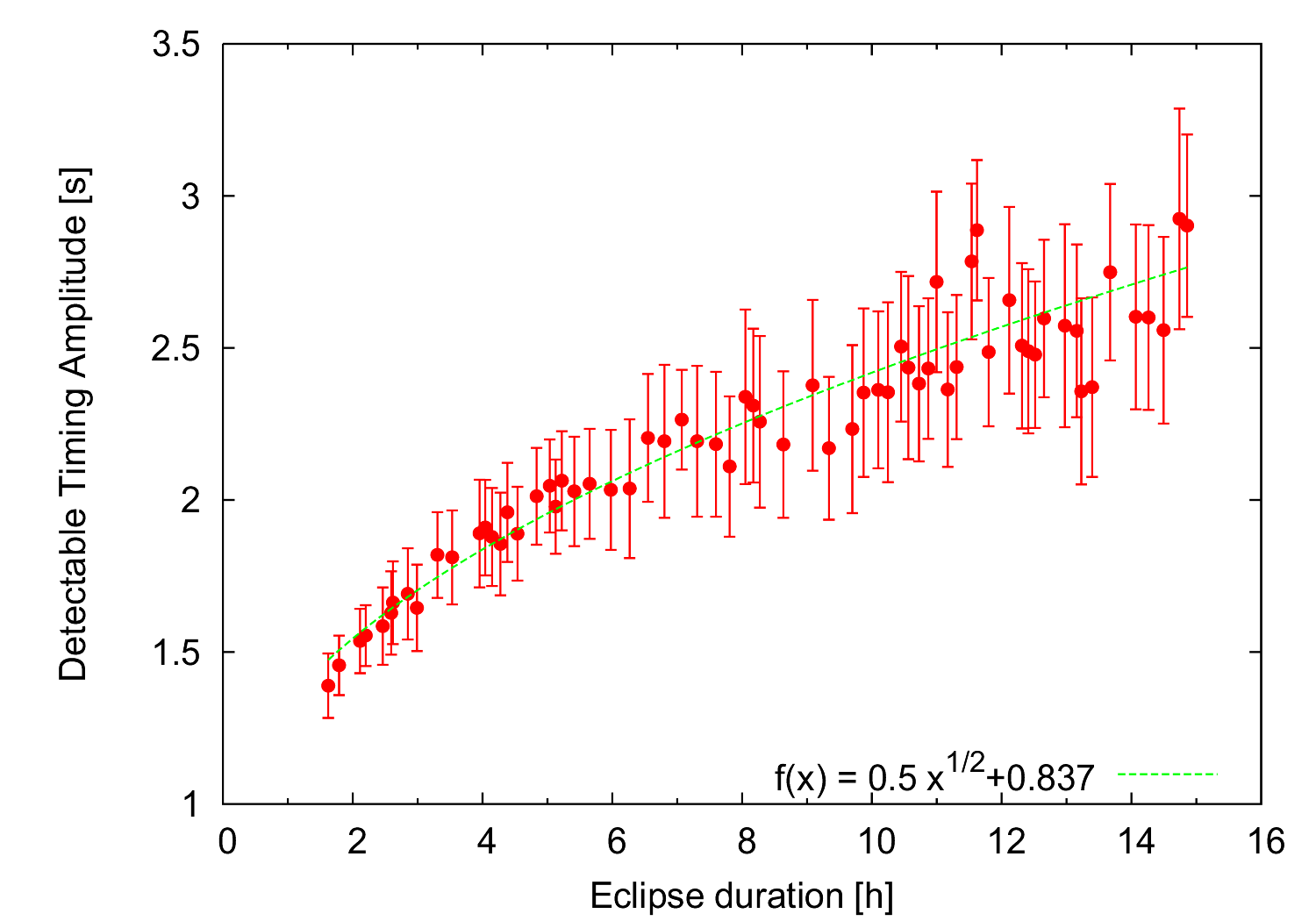}
\caption{Duration of an eclipse and DTA.
The solid line represents the analytic approximation by 
\citealt{doyle04a}.}
\label{fig:12_eclDu}
\end{figure}

\begin{figure}
\includegraphics[width=\columnwidth]{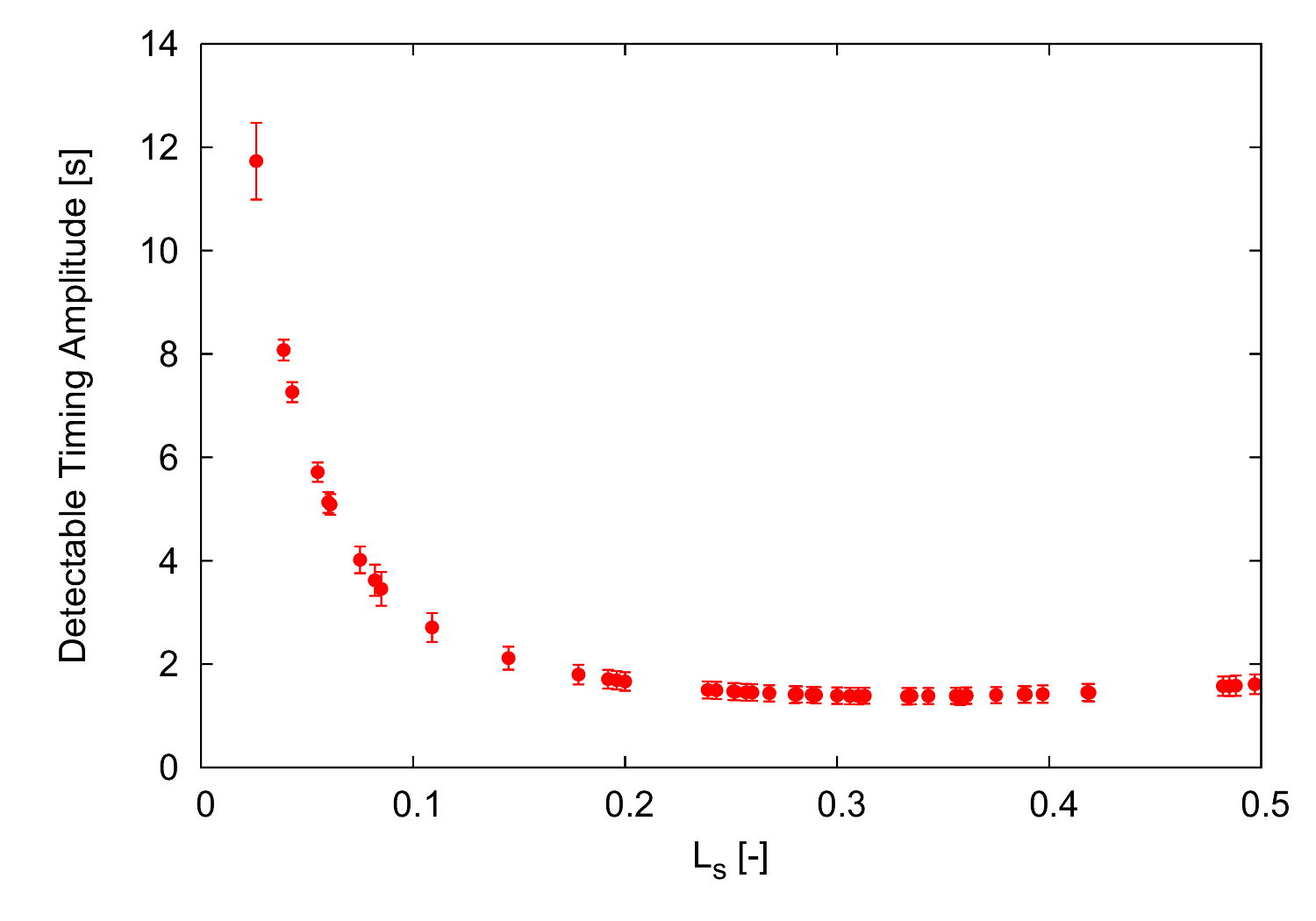}
\caption{Impact of the second star's luminosity $L_S$ 
in units of the primary star's luminosity on the detectable
timing amplitude.}
\label{fig:09_efLs}
\end{figure}

\begin{figure}
\includegraphics[width=\columnwidth]{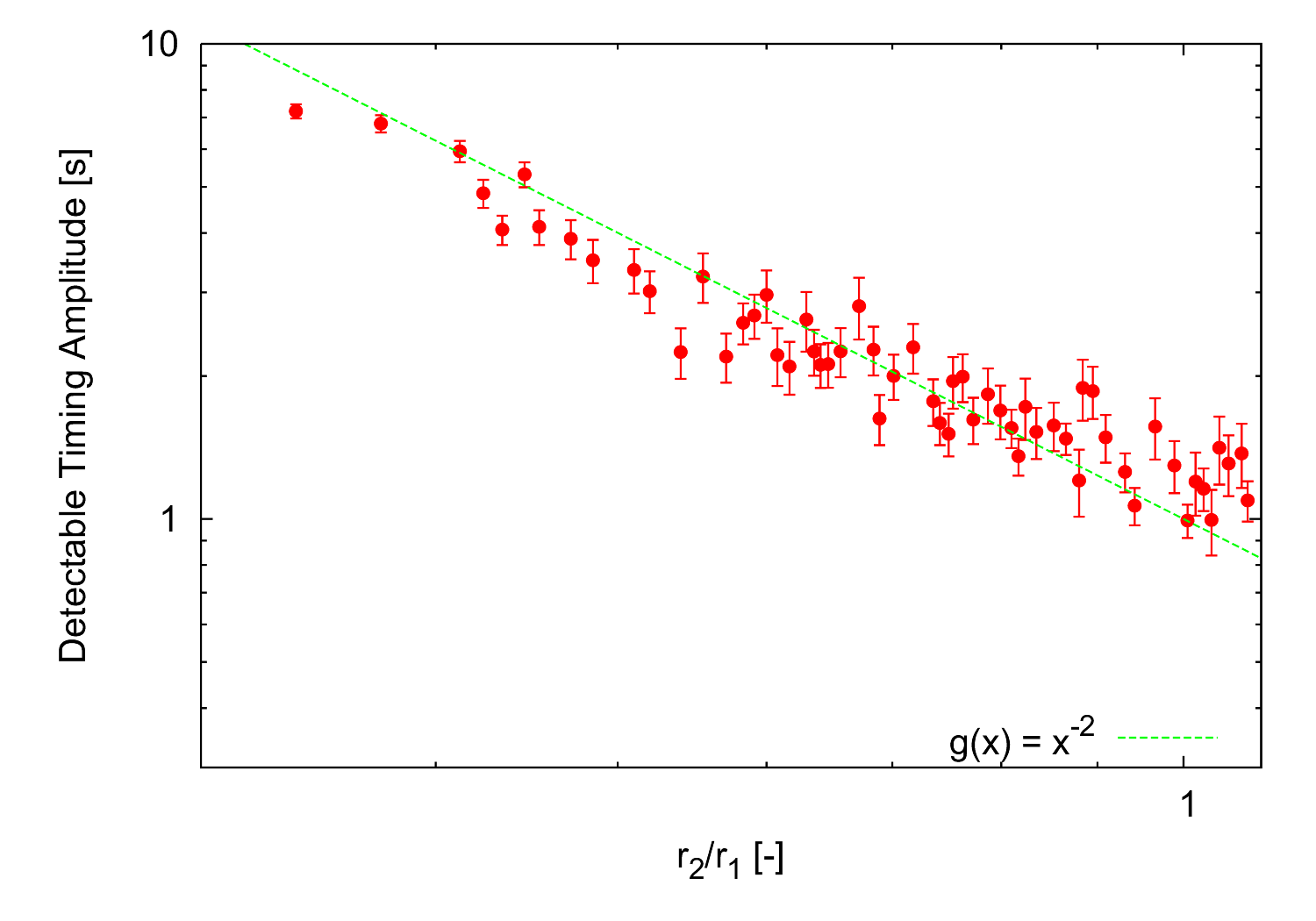}
\caption{Detectable timing amplitude as a function of
the ratio of radii of the secondary and primary star
($r_2/r_1$). In these simulations the parameters for
the ground based case were used.}
\label{fig:04_RadiusDSE}
\end{figure}

Let us note that \cite{doyle04a} derived an equation
allowing one to estimate an eclipse timing precision assuming
that the eclipse has a simple triangular shape
\begin{equation}
\label{eq08:doyleDeeg}
\delta_{t_0}\approx \delta_{L}\frac{T_{ec}}{2\Delta L \sqrt{N}}
\end{equation}
where $T_{ec}$ is the duration of an eclipse, $N$ is the number of
photometric measurements taken during $T_{ec}$ and 
$\Delta L$ is the relative depth of the eclipse. Our
simulations prove that this is a good approximation. This can be seen 
in Figure \ref{fig:12_eclDu} where one should note that after introducing 
the integration time $T_{int}$, we have $N=T_{ec}/T_{int}$ and
the equation \ref{eq08:doyleDeeg} can be rewritten as
\begin{equation}
\label{eq08b:doyleDeeg}
\delta_{t_0}\approx \delta_{L}\frac{\sqrt{T_{ec}T_{int}}}{2\Delta L}
\end{equation}
In the above,  $T_{int}$ is constant, the DTA is 
approximately equal to $\delta_{t_0}$ and the square 
root relation between $\delta_{t_0}$ and $T_{ec}$ is visible. 

\begin{figure}
\includegraphics[width=\columnwidth]{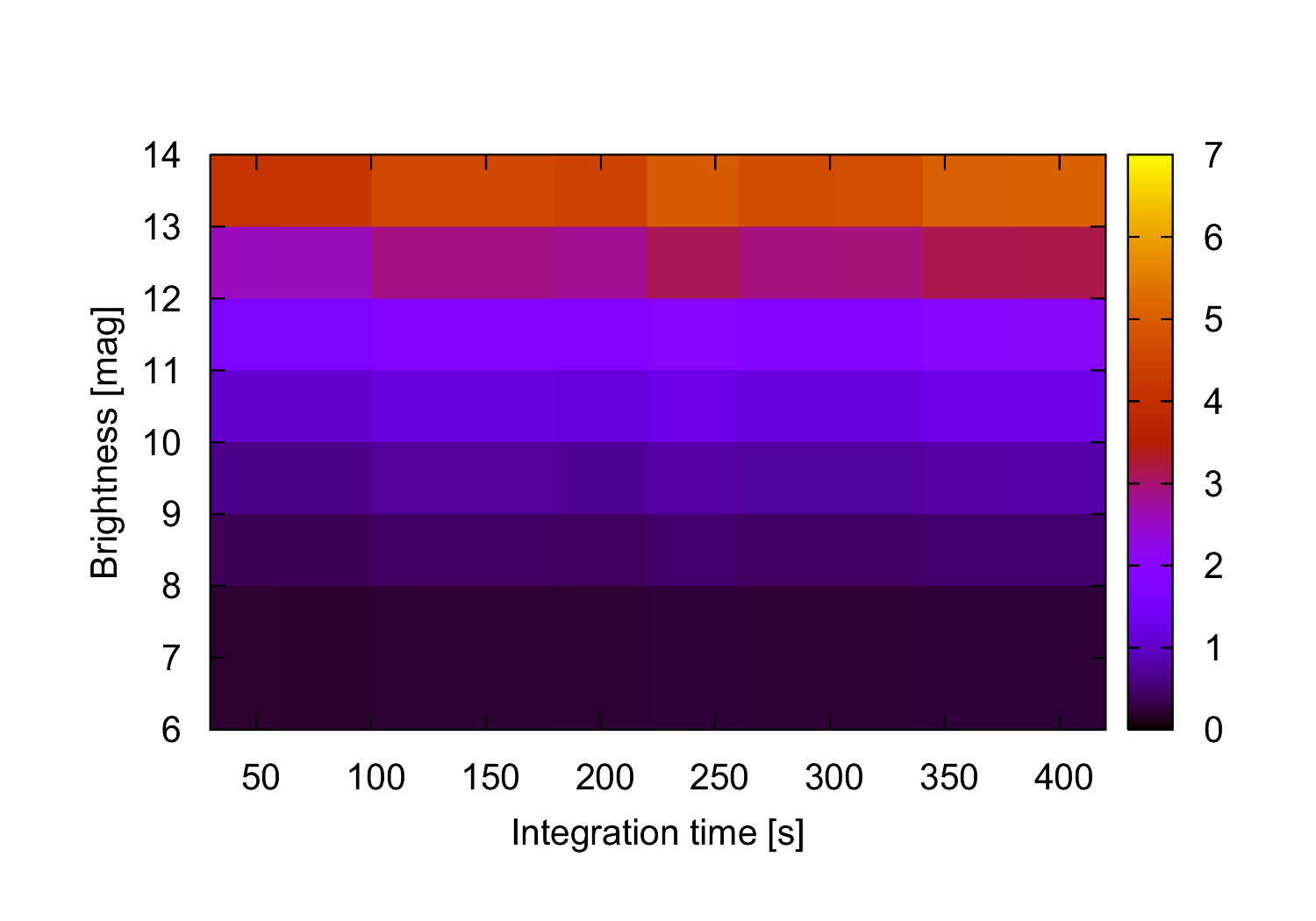}
\caption{DTA as a function of the integration time and brightness
of the target for a ground-based effort.}
\label{fig:08_GroundDtaPN}
\end{figure}

\begin{figure}
\includegraphics[width=\columnwidth]{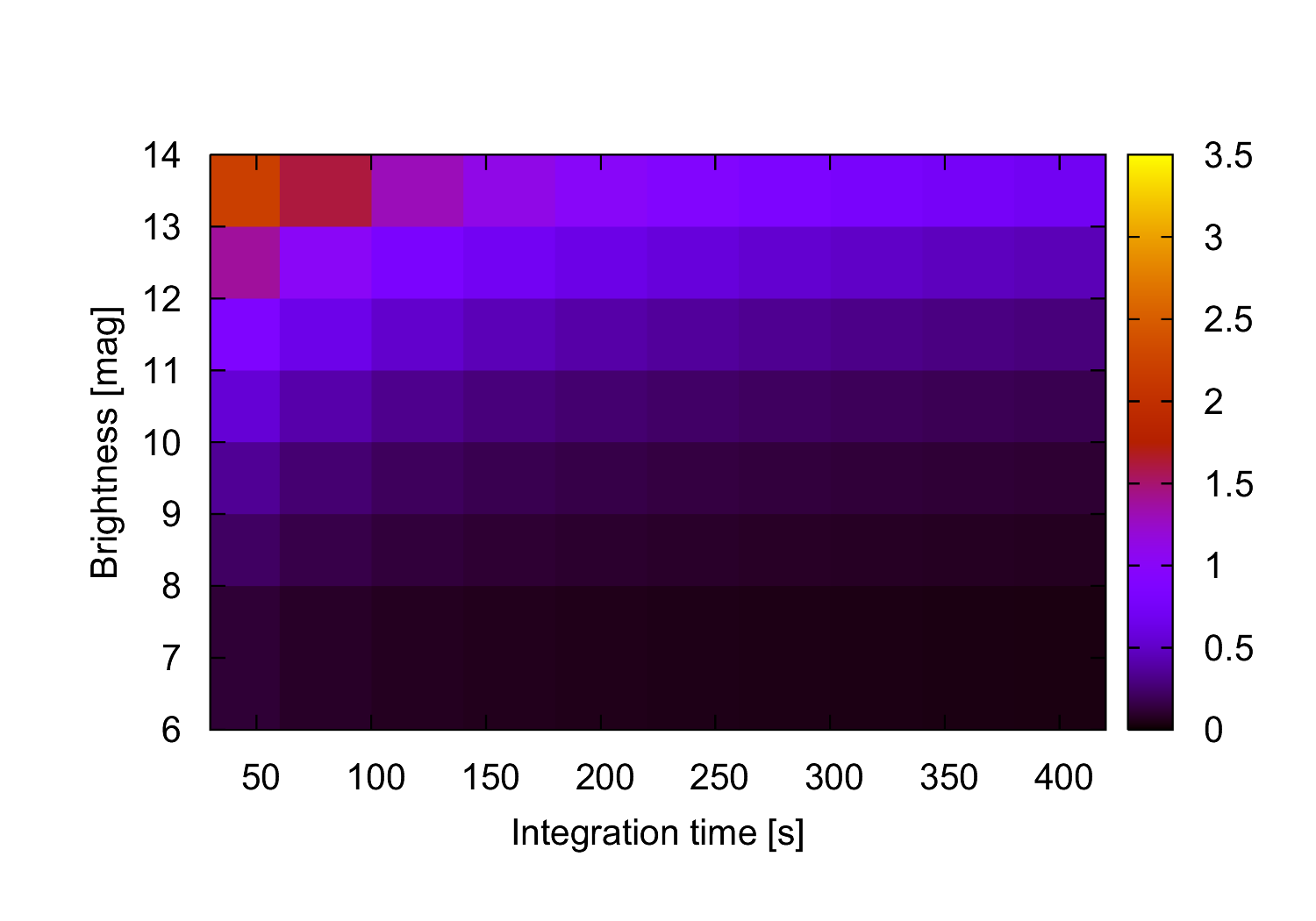}
\caption{Photometric precision in mmag as a function of the integration time and
brightness of the target for a ground-based effort.}
\label{fig:09_GroundSigPN}
\end{figure}

We conclude the simulations with two representative figures for
a ground base survey based on a 0.5-m telescope. Figure~\ref{fig:08_GroundDtaPN}
shows DTA and Figure~\ref{fig:09_GroundSigPN}
a typical photometric error due to the photon noise for our test scenario. 


\section{Conclusions}
\label{sec6}
\begin{figure}
\includegraphics[width=\columnwidth]{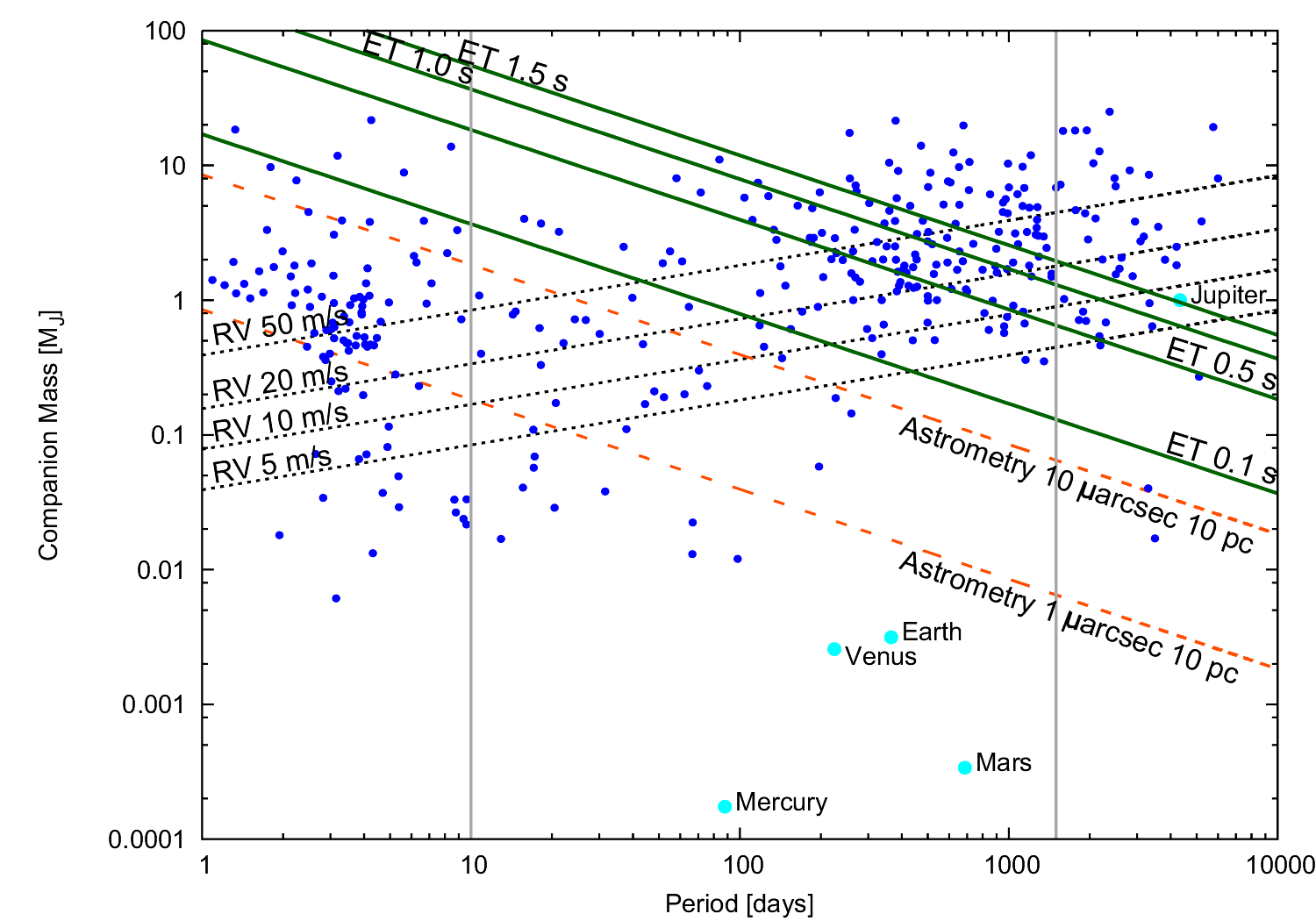}
\caption{Discovery space for circumbinary planets around a binary star
composed of two Sun-like stars. Known exoplanets are marked with dots. The 
two vertical lines correspond to the shortest stable orbit for this case
\citep{dvorak89,holman} and 4 years.}
\label{fig:13_limits}
\end{figure}

In order to detect giant circumbinary planets around eclipsing stars 
a timing precision of the order of 0.1-1 seconds is necessary.
The Kepler and CoRoT missions are capable of providing photometric precision
sufficient to reach such a timing precision. However, in both cases 
what makes the detections challenging is a predefined target pool.
In the case of CoRoT typical targets are quite faint and the duration
of an observing window is only 150 days which effectively
limits the detection capabilities to brown dwarfs. In the case
of Kepler, the target pool puts an upper limit of potentially
detectable circumbinary gas giants at about 40 in the best case
scenario. This number does not take into account the orbital
and physical parameters of circumbinary planets (like e.g. masses). 
Hence, the more realistic upper limit is expected to be up to
several times lower.
Nevertheless, both missions still may deliver us a detection
of a circumbinary planet via eclipse timing.
It seems that the best strategy
to detect circumbinary planets around eclipsing binary
stars is by carrying out a ground based survey for which
targets can be carefully preselected. Such survey
would have to employ several 0.5-m class telescopes
to be efficient. As the survey would typically focus
on the shorter period detached eclipsing binary stars
(see Figure~10), it would target a different set
of targets than a radial velocity based survey.

In Figure \ref{fig:13_limits} we compare planet detection
capabilities of the radial velocity, astrometry and eclipse 
timing. Eclipse timing which is essentially a 1-d astrometric
measurement is complementary to the radial velocity
technique and as we have demonstrated one is able to achieve
a timing precision sufficient to detect giant planets.
Finally, let us note that two circumbinary planets around an eclipsing binary 
HW Vir \citep {lee}, a circumbinary brown dwarf around an eclipsing binary 
HS0705+6700 \citep{Q:09::} and a giant planet around an eclipsing
polar DP Leo \citep{{Q:10::}} were claimed to be detected by means of
eclipse timing. However, it is hard to judge if these cases
of timing variation are really caused by substellar companions
and not an unknown quasi-periodic phenomenon. Nevertheless 
this is yet another proof that eclipse timing is becoming a useful 
tool for detecting subtle timing variations.

\section*{Acknowledgments}
This work is supported by the Foundation for Polish Science 
through a FOCUS grant, by the Polish Ministry of Science and 
Higher Education through grant 
No.~N203~005~32/0449, and by the European Social Fund and the national
budget of the Republic of Poland within the framework of the Integrated
Regional Operational Programme, Measure 2.6. Regional innovation
strategies and transfer of knowledge - an individual project of the
Kuyavian-Pomeranian Voivodship "Scholarships for Ph.D. students
2008/2009 - IROP"

\bsp

\label{lastpage}


\begin{thebibliography}{}
\bibitem[\protect\citeauthoryear{Alonso et al.}{2008}]{alonso}
Alonso R., Auvergne M., Baglin A., Ollivier M., Moutou C., Rouan D., Deeg H. J., Aigrain S. et al., 2008, A\&A, 482, L21 

\bibitem[\protect\citeauthoryear{Auvergne et al.}{2009}]{auvergne} Auvergne M., Bodin 
P., Boisnard L., Buey J. -T., Chaintreuil S., Epstein G., Jouret M., Lam-Trong T. et al., 2009, A\&A, 506, 411

\bibitem[\protect\citeauthoryear{Costes et al.}{2004}]{costes04c} Costes V., Bodin P., Levacher P., Auvergne M., 2004, 5th International Conference on Space Optics, 554, 281 

\bibitem[\protect\citeauthoryear{Cumming}{2004}]{Cum:04::} 
Cumming A., 2004, MNRAS, 354, 1165 

\bibitem[\protect\citeauthoryear{Deeg et al.}{2008}]{deeg08a} Deeg H. J., Oca{\~n}a  B., Kozhevnikov V. P., Charbonneau D., O'Donovan F. T., Doyle L. R., 2008, A\&A, 480, 563

\bibitem[\protect\citeauthoryear{Doyle \& Deeg}{2004}]{doyle04a} Doyle L. R., Deeg H. J., 2004, Bioastronomy 2002: Life Among the Stars, 213, 80 

\bibitem[\protect\citeauthoryear{Dvorak}{1984}]{dvorak} Dvorak R., 1984, Celestial Mechanics, 34, 369 

\bibitem[\protect\citeauthoryear{Dvorak}{1989}]{dvorak89} Dvorak R., Froeschle C., Froeschle C., 1989, A\&A, 226, 335 

\bibitem[\protect\citeauthoryear{Garrido \& Deeg}{2006}]{garrido06} Garrido R., Deeg H. J., 2006, Lecture Notes and Essays in Astrophysics, 2, 27 

\bibitem[\protect\citeauthoryear{Holman \& Wiegert}{1999}]{holman} Holman M. J., Wiegert P. A., 1999, AJ, 117, 621 

\bibitem[\protect\citeauthoryear{Koch et al.}{2004}]{koch04b}
Koch D. G., Borucki W., Dunham E., Geary J., Gilliland R., Jenkins J., Latham D., Bachtell E. et al., 2004, Proc. SPIE, 5487, 1491 

\bibitem[\protect\citeauthoryear{Lee et al.}{2009}]{lee} Lee J. W., Kim S.-L., 
Kim C.-H., Koch R. H., Lee C.-U., Kim H.-I., Park J.-H., 2009, AJ, 137, 3181 

\bibitem[\protect\citeauthoryear{Milotti}{2006}]{milotti} Milotti E., 2006, Computer 
Physics Communications, 175, 212 

\bibitem[\protect\citeauthoryear{Milotti}{2007}]{milotti07} Milotti E., 2007, Phys. Rev, 75, 
011120 

\bibitem[\protect\citeauthoryear{Mor{\'e}, Garbow \& Hillstrom}{Mor{\'e} et al.}{1980}]{more} Mor{\'e} J. J., Garbow B. S., Hillstrom K. E., 1980, User Guide for MINPACK-1, Argonne National Laboratory Report ANL-80-74

\bibitem[\protect\citeauthoryear{Mor{\'e} et al.}{1984}]{more84a} Mor{\'e} J. J., Sorensen D. C., Hillstrom K. E., Garbow, B. S., 1984, The MINPACK Project, in Sources and Development of Mathematical Software, W. J. Cowell, ed., 88-111

\bibitem[\protect\citeauthoryear{Muterspaugh et al.}{2007}]{muterspaugh07a} Muterspaugh M. W., Konacki M., Lane B. F., Pfahl E., 2007, preprint (arXiv:astro-ph/0705.3072v1) 

\bibitem[\protect\citeauthoryear{Nelson \& Davis}{1972}]{nelson} Nelson B., Davis W. D., 1972, ApJ, 174, 617 

\bibitem[\protect\citeauthoryear{Ofir, Deeg \& Lacy}{Ofir et al.}{2009}]{ofir09b} Ofir A., Deeg H. J., Lacy C. H. S., 2009, A\&A, 506, 445

\bibitem[\protect\citeauthoryear{Paczy{\'n}ski et al.}{2006}]{paczynski06a} Paczy{\'n}ski B., Szczygie{\l} D. M., Pilecki B., Pojma{\'n}ski G., 2006, MNRAS, 368, 1311 

\bibitem[\protect\citeauthoryear{Qian et al.}{2010}]{Q:10::} 
Qian S.-B., Liao W.-P., Zhu L.-Y., Dai Z.-B., 2010, ApJ, 708, L66 

\bibitem[\protect\citeauthoryear{Qian et al.}{2009}]{Q:09::} 
Qian S.-B., Dai Z.-B., Liao W.-P., Zhu L.-Y., Liu L., Zhao E.~G., 2009, 
ApJ, 706, L96 

\end{thebibliography}
\end{document}